\documentclass[a4paper,11pt]{article}    
\pdfoutput=1
\usepackage[OT1]{fontenc}
\usepackage{graphicx} 
\usepackage{bbold}
\usepackage{mathtools}
\usepackage{amsmath}
\numberwithin{equation}{section} %equation numbering
\usepackage{amsfonts}
\usepackage{amssymb}
\usepackage{amsbsy}
\usepackage{amsfonts}
\usepackage{physics}
\usepackage{bbold}
\usepackage{slashed} 
\usepackage{color}
\usepackage[table]{xcolor}
\usepackage{tablestyles}
\usepackage{tabularx}
\usepackage{hyperref}
\usepackage{bm}% bold math
\usepackage{orcidlink}
\usepackage{bbding}
\usepackage{cite}  %  to make successive citations hyphened
\usepackage{mathrsfs}
\usepackage{framed}
\usepackage[font=footnotesize,labelfont=bf]{caption}
\def\eq#1{{Eq.~(\ref{#1})}}
\def\eqs#1#2{{Eqs.~(\ref{#1})--(\ref{#2})}}

\def\abs#1{\left| #1\right|}

\def\Tr{\mbox{Tr}\,}

\newcommand{\di}{\mbox{d}}

\def\di{\mbox{d}}

\colorlet{grayline}{gray!70}
\definecolor{blueline}{rgb}{0,0.27,0.55}
\definecolor{DarkGray}{gray}{0.4}
\definecolor{Gray}{gray}{0.6}
\definecolor{oucrimsonred}{rgb}{0.6, 0.0, 0.0}
\definecolor{persianblue}{rgb}{0.11, 0.22, 0.73}
\definecolor{forestgreen}{rgb}{0.13,0.35,0.13}
 \hypersetup{colorlinks, citecolor=forestgreen, linkcolor=forestgreen, urlcolor=forestgreen}
\newcommand{\be}{\begin{equation}}
\newcommand{\ee}{\end{equation}}
\newcommand{\bea}{\begin{eqnarray}}
\newcommand{\eea}{\end{eqnarray}}

\newcommand{\CC}{\operatorname{C}}
\newcommand{\BB}{\operatorname{B}}

\newcommand*\xbar[1]{%
  \hbox{\;%
    \vbox{%
      \hrule height 0.5pt % The actual bar
      \kern0.5ex%         % Distance between bar and symbol
      \hbox{%
        \kern-0.25em%      % Shortening on the left side
        \ensuremath{#1}%
        \kern-0.07em%      % Shortening on the right side
      }%
    }%
  }%
} 
\newcommand{\com}[1]{}
\newcommand{\gsim}{\lower.7ex\hbox{$\;\stackrel{\textstyle>}{\sim}\;$}}
\newcommand{\lsim}{\lower.7ex\hbox{$\;\stackrel{\textstyle<}{\sim}\;$}} 

\newcommand{\bc}{\begin{center}}
\newcommand{\ec}{\end{center}}

\font\beeg=cmr17 scaled 1800
\newbox\ibox
\def\versal#1{\setbox\ibox=\hbox{{\beeg #1}~}%
	    \noindent\global\hangindent=\wd\ibox\global\hangafter-2%
	    \sc\smash{\llap {\lower 14pt \box\ibox}}}
\begin{document}
\onecolumn
\thispagestyle{empty}
\bc
{ \Large \color{oucrimsonred} \textbf{ 
The trace distance between density matrices,\\[0.4em] a nifty tool in new-physics searches}}

\vspace*{1.5cm}
{\color{DarkGray}
  {\bf M. Fabbrichesi$^{a \, \orcidlink{0000-0003-1937-3854}}$},
  {\bf M. Low$^{b\, \orcidlink{0000-0002-2809-477X}}$,} and 
  {\bf L.~Marzola$^{c,d \, \orcidlink{0000-0003-2045-1100}}$}
}\\
\vspace{0.5cm}
{\small 
{\it  \color{DarkGray} (a)
INFN, Sezione di Trieste, Via Valerio 2, I-34127 Trieste, Italy}
  \\[1mm]  
  {\it \color{DarkGray}
(b) Pittsburgh Particle Physics, Astrophysics, and Cosmology Center,\\Department of Physics and Astronomy, University of Pittsburgh, Pittsburgh, USA
}
  \\[1mm]  
  {\it \color{DarkGray}
(c) Laboratory of High-Energy and Computational Physics, NICPB, R\"avala pst 10, \\ 10143 Tallinn, Estonia}
\\[1mm]  
  {\it \color{DarkGray}
(d) Institute of Computer Science, University of Tartu, Narva mnt 18, \\ 51009 Tartu, Estonia}
}
\ec

 \vskip0.5cm
\bc
{\color{DarkGray}
\rule{0.7\textwidth}{0.5pt}}
\ec
\vskip1cm
\bc
{\bf ABSTRACT}
\ec
\noindent
Quantum information methods have been brought to bear on high-energy physics, including the study of entanglement and Bell nonlocality in collider experiments. Quantum information observables have also been employed to constrain possible new physics effects. We improve on this point by introducing quantum information tools routinely used to compare quantum states: the trace distance and the fidelity. We find that the former outperforms other quantum information observables considered in the literature and, together with the cross section, yields the strongest bounds on possible departures from the Standard Model. The power of the proposed methodology is demonstrated with three examples of new physics searches. The first concerns the chromomagnetic dipole moment of the top quark and yields the first bound computed by means of quantum tomography and actual experimental data. The other two examples use Monte Carlo simulations and set the projected limits on the anomalous couplings of the $\tau$ leptons at Belle and at a future collider---which is taken to be LEP3.  For these new physics searches we also compare the sensitivity of the trace distance to those of other quantum information quantities like concurrence, magic, and the fidelity distance. In passing, we provide the first determinations  of magic in colliders data by analyzing the top-quark pair production at the LHC and the charmonium decays. The significance is well above the $5\sigma$ level in both the cases.

\vskip 2cm
\bc 
{\color{DarkGray} %\vbox{$\bowtie$} 
%\SixFlowerPetalDotted 
\SquareShadowBottomRight
}
\ec

%%%%%%%%%%%%%%%%%%%%%%%%%%%%%%%%%%%%%%%%%%%%%%%%%%%%%%%%%%%%%%%%%%
\newpage
%%%%%%%%%%%%%%%%%%%%%%%%%%%%%%%%%%%%%%%%%%%%%%%%%%%%%%%%%

\section{Introduction\label{sec:intro}} 
%%%%%%%%%%%%%%%%%%%%%%%%%%%%%%%%%%%%%%%%%%%%%%%%%%%%%%%%%

{\versal As we  probe the Standard Model (SM) }to define its limits, we are in constant need of better and better tools to find possible deviations that would signal the presence of new physics. The application of quantum information to high energy physics is the latest addition to this toolbox.  Quantum tomography is based on a set of complementary measurements that allow for the full reconstruction of the quantum state describing the system under investigation, resulting in a density operator. A typical example is the determination of the spin state characterizing an ensemble of particles produced in collider experiments, accessible for a large class of events with current detectors through the reconstruction of suitable angular distributions---see Ref.~\cite{Barr:2024djo} for a review and a detailed illustration of the method.

The spin state of an ensemble of spin-1/2 fermion pairs, for instance, is then encoded in a density operator of a bipartite qubit system.  The operator is represented by the $4\times4$ matrix
\begin{equation}
	\label{eq:rho_deco}
		\rho = \frac{1}{4} \qty[
		\mathbb{1}\otimes\mathbb{1} 
		+ 
		\sum_{i=1}^3 \BB_i^+ \, \qty(\sigma_i \otimes \mathbb{1}) 
		+ 
		\sum_{j=1}^3 \BB_j^- \, \qty(\mathbb{1} \otimes \sigma_j )
		+
		\sum_{i,j=1}^3 \CC_{ij} \, \qty(\sigma_i \otimes \sigma_j)
		],
\end{equation}
where we have denoted with $\sigma_i$ ($i=1,2,3$) the Pauli matrices, with $\mathbb{1}$ the unit matrix of dimension 2, and with $\otimes$ their Kronecker products. The full ensemble is characterized by 15 Fano coefficients: 6 give the polarizations $\BB^\pm$---the averages of the spin vectors---while the remaining 9 are organized in the matrix $\CC$ that accounts for the spin correlations. The coefficients can be formally computed as   
    \be
		\BB_i^+ = \Tr[\rho\qty(\sigma_i \otimes \mathbb{1}) ]\,,\quad
		\BB_i^- = \Tr[\rho\qty( \mathbb{1} \otimes \sigma_i) ]\,,\quad
		\CC_{ij} = \Tr[\rho\qty(\sigma_i \otimes \sigma_j)]\,,
        \label{eq:FanoC}
	\ee
and implicitly refer to the frame of reference used to describe the orientation of the spin vectors in space. As mentioned before, these coefficients can be experimentally reconstructed by analyzing the angular distribution of suitable decay products of the fermions, which reveal the orientation of the spin vectors of the progenitor particles. The coefficients can also be analytically computed from the amplitude of the underlying pair-production process within any particle physics model, thereby allowing for its experimental test. 

By means of quantum tomography, entanglement (for a definition see, for instance, Ref.~\cite{Horodecki:2009zz}) has been argued to be present in top-quark pair production~\cite{Afik:2020onf}  at the Large Hadron Collider (LHC) and the violation of Bell's inequality~\cite{Bell:1964} (a signature of quantum nonlocality) has been shown to be experimentally accessible in the same system~\cite{Fabbrichesi:2021npl}.  Both entanglement and Bell inequality violation have been deemed potentially testable in the decay of the Higgs boson into charged~\cite{Barr:2021zcp} or neutral~\cite{Aguilar-Saavedra:2022wam} gauge bosons, and in $\tau$-lepton pairs produced at Belle~\cite{Ehataht:2023zzt}. Entanglement and Bell inequality violation have been found in reinterpretations of the LHCb and Belle data in the decays of the $B$ mesons, with significances well in excess of 5$\sigma$~\cite{Fabbrichesi:2023idl}, as well as in reinterpretations of BESIII data in several charmonium decays~\cite{Fabbrichesi:2024rec}.  The ATLAS and CMS collaborations have directly found~\cite{ATLAS:2023fsd,CMS:2024pts} that entanglement is present in top-quark pairs produced near threshold at the LHC.  The CMS collaboration has recently published an analysis of the same process at higher invariant masses as well~\cite{CMS:2024zkc}.

Quantum tomography and entanglement are currently also being used in setting bounds on new physics in several processes: top-quark pair production~\cite{Aoude:2022imd,Fabbrichesi:2022ovb,Maltoni:2024tul}, Higgs physics~\cite{Altakach:2022ywa,Fabbrichesi:2023jep,Bernal:2023ruk}, diboson production~\cite{Aoude:2023hxv} and $\tau$-lepton pair production at Belle~\cite{Fabbrichesi:2024xtq} and at the FCC-ee~\cite{Fabbrichesi:2024wcd}. 

The density matrix of a system is the gateway towards achieving its full characterization. Knowledge of it gives access to a plethora of observables typically employed only within the quantum information field, where entanglement, discord, steering, magic, and Bell inequality violation are of importance for cryptographic and quantum computing protocols. These phenomena can also be investigated at collider experiments, in energy ranges and interactions well outside the reach of the typical quantum information tests~\cite{Afik:2022dgh,White:2024nuc,Han:2024ugl}. Although these observables provide new means to constrain physics beyond the SM, it is the density operator that encapsulates most, if not all, of the available information pertaining to the system under consideration.

In order to fully capitalize on the possibility offered by quantum tomography, in this paper we introduce into high-energy physics two quantities that are routinely used in quantum information to compare two quantum states: the \textit{trace distance} and the \textit{fidelity}. Differently from other observables, trace distance and fidelity utilize the full density matrices of the two target quantum systems and consequently offer, once supplemented with the cross section, the means to best constrain parameters that could source potential discrepancies. 

In the following, after introducing these quantities and briefly reviewing their properties, we demonstrate their use by studying the internal structure of the top quark in QCD by means of recent LHC data. We then employ them to constrain the anomalous couplings of the $\tau$ lepton to gauge bosons, referring to two benchmark cases tailored to the LEP3 future collider and the ongoing Belle experiment by means of Monte Carlo simulations. For the latter setup we also sketch the possibilities offered by the spin formalism within quantum tomography. We conclude the paper with a comparison of the power in constraining new physics of different quantum information observables.

%%%%%%%%%%%%%%%%%%%%%%%%%%%%%%%%%%%%%%%%%%%%%%%%%%%%%%%%%
\section{Distance and similarity of quantum states\label{sec:distances}}
%%%%%%%%%%%%%%%%%%%%%%%%%%%%%%%%%%%%%%%%%%%%%%%%%%%%%%%%%

{\versal The trace distance and fidelity} are used to quantify the difference and the similarity between two quantum states, in the same way as their classical counterparts quantify the difference and similarity of two probability distributions.

The trace distance between two density matrices $\rho$ and $\varsigma$ is defined as~\cite{Nielsen:2012yss}
\be
\mathscr{D}^T\qty(\rho, \varsigma)=\dfrac{1}{2} \, \Tr \sqrt{(\rho - \varsigma)^\dag (\rho - \varsigma)} \geq0 \, ,\label{D}
\ee
generalizing the Kolmogorov distance used for probabilities distributions. The trace distance is a metric on the space of density operators and remains invariant under unitary transformations: $\mathscr{D}^T(\rho, \varsigma)=\mathscr{D}^T(U\,\rho\, U^\dag, U\, \varsigma \,U^\dag)$, with $U$ a unitary matrix. To see the effect of the trace distance explicitly, consider the simple case in which both $\rho$ and $\varsigma$ describe, each, a qubit. Then, given
\begin{equation}
    \rho = \frac{1}{2}\Big[\mathbb{1}+\vec{r}\cdot \vec\sigma \Big]\,,\qquad \varsigma = \frac{1}{2}\Big[\mathbb{1}+\vec{s}\cdot \vec\sigma \Big] \, ,
\end{equation}
with $\vec \sigma$ being the vector of Pauli matrices, it follows
\begin{equation}
    \mathscr{D}^T(\rho, \varsigma) = \frac{\norm{\vec{r} - \vec{s}}}{2} \,, 
\end{equation}
whereas if $\qty[\rho ,\sigma]=0$, the trace distance recovers the classical expression with probability distributions given by the eigenvalues of the density matrices. 

Contrary to the trace distance, the fidelity  
\be
\mathscr{F}\qty(\rho, \varsigma) = \Tr \sqrt{\sqrt\rho \,\varsigma \,\sqrt\rho} \, ,
\ee
is a similarity measure. Although not manifestly, it holds $\mathscr{F}\qty(\rho, \varsigma)$ = $\mathscr{F}\qty(\varsigma, \rho)$ and the fidelity is bounded to $0\leq \mathscr{F}(\rho, \varsigma)\leq 1$, with the lower bound being saturated if and only if the density matrices have orthogonal supports, and the upper bound if and only if $\rho= \varsigma$. Analogously to the trace distance, the fidelity is invariant under unitary transformations and recovers its classical counterpart when the density matrices commute. That the fidelity is a measure of similarity can be seen from Uhlmann's theorem~\cite{Uhlmann:1976def}
\begin{equation}
    \mathscr{F}\qty(\rho, \varsigma) = \max_{\ket{r}, \ket{s}}\abs{\braket{r}{s}}\,,
\end{equation}
in which $\ket{r}$ and $\ket{s}$ are purifications\footnote{Given a density operator $\rho$ defined over the Hilbert space $\mathcal{H}$, we define its purification as a bipartite pure state $\ket{\Psi}\in\mathcal{H}\otimes\mathcal{H}_R$ built with a fictitious reference system whose possible states span the Hilbert space $\mathcal{H}_R$. Then it holds $\rho = \Tr_{R}(\dyad{\Psi})$, where the partial trace is taken over the fictitious system.} of $\rho$ and $\varsigma$, respectively. The fidelity reduces to the overlap $ \mathscr{F}\qty(\dyad{r}, \dyad{s}) = \abs{\braket{r}{s}}$ if the two systems being compared are described by pure states $\ket{r}$ and $\ket{s}$.

Fidelity and trace distance are related via the Fuchs–van de Graaf inequalities~\cite{Fuchs:1997ss}
\begin{equation}
    1 -  \mathscr{F}\qty(\rho, \varsigma)  \leq  \mathscr{D}^T\qty(\rho, \varsigma) \leq  \mathscr{D}^F\qty(\rho, \varsigma)\,,  \label{inequality}
\end{equation}
where we introduced the fidelity distance~\cite{Gilchrist_2005}
\begin{equation}
    \mathscr{D}^{F}=\sqrt{1- \mathscr{F}^2} \,.\label{DF}
\end{equation}
The fidelity distance is also a metric on the space of density operators~\cite{Gilchrist_2005}.\footnote{The fidelity distance corresponds to the $C(\cdot,\cdot)$ metric of Ref.~\cite{Gilchrist_2005}.}
For pure states, we have $\mathscr{D}^T\qty(\dyad{r}, \dyad{s})=\mathscr{D}^F\qty(\dyad{r}, \dyad{s})$.  Other fidelity-based metrics are the Bures metric, $\sqrt{2- 2\mathscr{F}}$, and the angle metric, $\arccos{\mathscr{F}}$.  A detailed comparison of these metrics is left for future work.

If we take for $\rho$ the density matrix for a Standard Model process and for $\varsigma$ the one obtained by including new physics effects, the trace distance and the fidelity can be used to constrain the latter. 

The distance measures  in \eqs{D}{DF}  can be written explicitly for Bell-diagonal states in a simple form.  In terms of the Fano coefficients of \eq{eq:rho_deco} Bell-diagonal (BD) states have the property that $B^- = B^+ = (0,0,0)$.  Consequently, the spin correlation matrix can be diagonalized and the state can be fully characterized by the three diagonal entries of the spin correlation matrix.  Consider the following two states:
\begin{equation}
\rho_{\rm BD} = \frac{1}{4} \qty[ \mathbb{1} + \sum_{i=1}^3 R_{ii} (\sigma_i \otimes \sigma_i)  ],
\qquad
\varsigma_{\rm BD} = \frac{1}{4} \qty[ \mathbb{1} + \sum_{i=1}^3 S_{ii} (\sigma_i \otimes \sigma_i)  ].
\end{equation}
The trace distance between these states is
\begin{equation}
\begin{aligned}
\mathscr{D}^T (\rho_{\rm BD}, \varsigma_{\rm BD})
& =
\frac{1}{8} \bigg(
| (R_{11} - S_{11}) + (R_{22} - S_{22}) + (R_{33} - S_{33}) | \\
& \qquad + | (R_{11} - S_{11}) + (R_{22} - S_{22}) - (R_{33} - S_{33}) | \\
& \qquad + | (R_{11} - S_{11}) - (R_{22} - S_{22}) + (R_{33} - S_{33}) | \\
& \qquad + | (R_{11} - S_{11}) - (R_{22} - S_{22}) - (R_{33} - S_{33}) | 
\bigg).
\end{aligned}
\end{equation}
The fidelity distance between these states is
\begin{equation}
\begin{aligned}
\mathscr{D}^F (\rho_{\rm BD}, \varsigma_{\rm BD})
& = \bigg[ 1 - \frac{1}{16} \bigg( 
\sqrt{(1-R_{11}-R_{22}-R_{33})(1-S_{11}-S_{22}-S_{33})}   \\
& \qquad + \sqrt{(1-R_{11}+R_{22}+R_{33})(1-S_{11}+S_{22}+S_{33})} \\
& \qquad + \sqrt{(1+R_{11}-R_{22}+R_{33})(1+S_{11}-S_{22}+S_{33})} \\
& \qquad + \sqrt{(1+R_{11}+R_{22}-R_{33})(1+S_{11}+S_{22}-S_{33})}  \bigg)^2 \bigg]^{1/2}.
\end{aligned}
\end{equation}

%%%%%%%%%%%%%%%%%%%%%%%%%%%%%%%%%%%%%%%%%%%%%%%%%%%%%%%%%
\subsection{Using the new tools}

The trace and fidelity distances can be used to analyze possible deviations from the SM result caused by new-physics (NP) parameters. For the simplest case of a single operator controlled by the parameter $\lambda$, the statistical $\chi^2$ test
\be
\chi^2(\lambda) = \left( \frac{\mathscr{D}^T\qty[\rho_{\text{\tiny NP}}(\lambda), \rho_{\text{SM}}]}{\sigma_{\mathscr{D}^T}}\right)^2 \leq (1.00)\; 3.84 \,,  \label{chi2}
\ee
written here for the trace distance, provides the bound on the parameter at the (68\%) 95\% confidence level (CL). The uncertainty $\sigma_{\mathscr{D}^T}$ encodes both experimental and theoretical errors that are relevant to the reconstruction of the SM density matrix $\rho_{\text{SM}}=\rho_{\text{\tiny NP}}\qty(\lambda=0)$. A likelihood test can be defined in a similar manner, but the $\chi^2$ test is sufficient for the purpose of demonstrating the use of the new tools.

Since any density matrix $\rho$ is normalized to $\Tr \rho =1$, information on the total number of events is not included in the $\chi^2$ test in \eq{chi2}. The test can then be extended to include a term that accounts for the cross section $\sigma$, obtaining
\be
\chi^2(\lambda) =  \left( \frac{\mathscr{D}^T\qty[\rho_{\text{\tiny NP}}(\lambda), \rho_{\text{SM}}]}{\sigma_{\mathscr{D}^T}}\right)^2 +  \left( 
\frac{\sigma_{\text{\tiny NP}}(\lambda)-\sigma_{\text{SM}}}{\sigma_{\sigma}}\right)^2 \leq (1.00)\; 3.84  \,,  \label{chi2last}
\ee
in which $\sigma_{\sigma}$ is the overall uncertainty of the corresponding cross section and $\sigma_{text{SM}} = \sigma_{\text{\tiny NP}}(\lambda=0)$. Instead of the cross section, in \eq{chi2last}, we could use an asymmetry based on the cross section data, for instance the forward-backward asymmetry, should this be more effective in constraining new physics effects.

In the examples above, and in most of the analyses presented in the the paper, we only use the trace distance because we find that it performs the best of all the distance measures we study.  While by \eq{inequality} the fidelity distance is numerically larger than the trace distance, the performance depends on the propagation of uncertainties on Fano coefficients to the distance itself.

An alternative to using distances is doing a $\chi^2$ test directly on the set of Fano coefficients.\footnote{CMS performed a $\chi^2$ test on a subset of the Fano coefficients in Ref.~\cite{CMS:2019nrx}.}  
In general, the relative performance between the trace distance and the direct $\chi^2$ test varies depending on the physics process and relative measurement uncertainties.  In the cases we study they perform similarly.
The trace distance is still preferable, however, because it has well-defined quantum informational properties.  For example, the trace distance is directly related to the probability that given a measurement of a state, that could be either $\rho$ or $\varsigma$, the state is correctly identified~\cite{Gilchrist_2005}.   Additionally, the trace distance allows us to disregard possible correlations among the Fano coefficients.

In the following sections we apply these quantum information operators to the study of the anomalous couplings of the top quark and the $\tau$ lepton.

%%%%%%%%%%%%%%%%%%%%%%%%%%%%%%%%%%%%%%%%%%%%%%%%%%%%%%%%%
\section{Top-quark chromomagnetic dipole moment}
%%%%%%%%%%%%%%%%%%%%%%%%%%%%%%%%%%%%%%%%%%%%%%%%%%%%%%%%%

{\versal As a first application} of the  proposed distance measures to a particle physics problem, we consider the presence of a chromomagnetic dipole term in the interaction Lagrangian of the top quark and the gluon: 
\be
\mathcal{L}\supset -i \, \mu_t \, \frac{g_s}{2\, m_t} \, \bar t \, \sigma^{\mu \nu} q_\nu  \, T^a\, t \, G_{\mu}^a\, , \label{eq:dipole}
\ee
in which $\sigma^{\mu\nu} =i/2 \,[\sigma^\mu,\, \sigma^\nu]$, $\sigma^\mu=(1,\vec \sigma)$, $T^a$ are the Gell-Mann matrices divided by a factor of two, and $q_\nu$ is the momentum of the gluon.
The Lagrangian in \eq{eq:dipole} gives rise to the effective vertices of one gluon or two gluons attached to the quark line.

In terms of the conventions used within the SM effective field theory, in which the chromomagnetic Lagrangian for the top quark is expressed in terms of $SU(2)_L\otimes U(1)_Y$ invariants, the vertex in \eq{eq:dipole} sources the operator
\be
\frac{c_{tG}}{\Lambda^2} \big( {\cal O}_{tG} +{\cal O}_{tG}^\dag \big) \quad \text{with}  \quad
 {\cal O}_{tG} =g_s \left(\bar Q_L \, \sigma^{\mu \nu} \, T^a\, t_R \right) \tilde{H}  G_{\mu\nu}^a \, ,
\ee
and establishes the correspondence 
\be
\mu_t = - \frac{\sqrt{2}\, m_t\, v}{\Lambda^2} c_{tG}\ ,
\ee
so that $c_{tG}/\Lambda^2 =0.1/[1 \text{TeV}]^2$ yields $\mu_t =-0.006$. Notice the change of sign. In the equation above,  $Q_L$ and $t_R$ stand for the $SU(2)_L$ doublet formed by left-handed top and bottom quark fields and for the right-handed top quark field, respectively. The symbol $\tilde{H}$ indicates, as usual, the dual of the $SU(2)_L$ Higgs doublet with a SM vacuum expectation value given by $\langle 0| \tilde{H} |0\rangle =(v/\sqrt{2},0)^T$ and $v\simeq246$ GeV.

The density matrix describing the spin state of $t\bar t$ pairs can be reconstructed from the spin correlation coefficients only, as the $\BB^\pm$ coefficients vanish due to $C$ and $P$ symmetries which QCD respects. The $\CC$ coefficients are given as
\be
\CC_{ij} [m_{t\bar t},\, \Theta,\, \mu_t]= \frac{L^{gg} (\tau)\, \tilde \CC_{ij}^{gg}[m_{t\bar t},\, \Theta,\, \mu_t]+L^{qq} (\tau)\, \tilde \CC_{ij}^{qq}[m_{t\bar t},\, \Theta,\, \mu_t]} {L^{gg}(\tau) \, \tilde A^{gg}[m_{t\bar t},\, \Theta,\, \mu_t]+L^{qq} (\tau)\, \tilde A^{qq}[m_{t\bar t},\, \Theta,\, \mu_t]} \, ,\label{pdf}
\ee
in terms of the coefficients $\tilde \CC_{ij}^{gg}$, $\tilde \CC_{ij}^{qq}$, $\tilde A^{gg}$, and $\tilde A^{qq}$  collected in the Appendix of Ref.~\cite{Fabbrichesi:2022ovb} both for the SM and for the new physics effects encoded in \eq{eq:dipole}. The combination of the two production channels in \eq{pdf}, $g+g\to t +\bar t$ and $q + \bar q \to t +\bar t$, is weighted by the respective parton luminosity functions
\be
L^{gg} (\tau)= \frac{2 \tau}{\sqrt{s}} \int_\tau^{1/\tau} \frac{\di z}{z} q_{g} (\tau z) q_{g} \left( \frac{\tau}{z}\right)\quad \text{and}\quad
L^{qq} (\tau)= \sum_{q=u,d,s}\frac{4 \tau}{\sqrt{s}} \int_\tau^{1/\tau} \frac{\di z}{z} q_{q} (\tau z) q_{\bar q} \left( \frac{\tau}{z}\right)\, ,
\ee
in which the functions $q_j(x)$ are the parton distribution functions (PDFs) for species $j$. For their evaluation we utilize the {\tt PDF4LHC21}~\cite{PDF4LHCWorkingGroup:2022cjn} set with $\sqrt{s}=13$ TeV and factorization scale $q_0=m_{t\bar t}$.

%The expression in \eq{pdf} must be expanded in the case of new physics to retain only terms linear in the new physics parameter. The linear approximation in the inclusion of the new physics effects used in Ref.~\cite{Fabbrichesi:2022ovb} is justified as long as the dipole operator, which scales with the energy of the process, is much smaller than the SM contribution. Quadratic corrections are expected to be very small. This is confirmed by the analysis in Ref.~\cite{Aoude:2022imd}, which shows that the effect of terms quadratic in the dipole operator are negligible in the kinematic region we are considering. 

\subsection{Consistency of the approximation}

The inclusion of new physics at linear order is justified as long as the dipole operator, which scales with the energy of the process, is much smaller than the SM contribution. A rough estimate can be obtained by taking the intermediate value  $m_{t\bar t} \simeq 1000$ GeV  and $\mu\simeq 0.01$, a typical value of the dipole. Hence, the new physics contribution is of the order
\be
\frac{m_{t\bar t} \, \mu}{ 2 m_t} \simeq 0.01, \label{est}
\ee
relative to the SM one. The linear approximation, which retains only single insertions of the dipole operator, is therefore well justified. Furthermore, for invariant masses $m_{t\bar{t}} <  O (1 \;{\rm TeV})$, the quadratic contributions of the magnetic-dipole operator are expected to be negligible as of order $O (m^2_{t\bar{t}}/\Lambda^2)$. Quadratic corrections are expected to be very small as confirmed by the leading order analysis in Ref.~\cite{Aoude:2022imd}. The next-to-leading order (NLO) computation appeared in Ref.~\cite{Severi:2022qjy} further confirms the result.
 
We also remark that the estimate in \eq{est} also shows that the SM effective field theory operator expansion is well defined, over the considered energy range, even for an operator such as the chromomagnetic dipole. Although this is expected to grow with energy, the corrections remain perturbative up to the scale of a few TeV, as involved in the present analysis.

A possible additional source of concern is posed by the magnitude of QCD NLO terms relative to new physics effects.
The one-loop QCD corrections give rise to a dipole operator with coefficient
\be
- \frac{\alpha_s}{\pi} \frac{m_t^2}{m_{t \bar t}^2} \log \frac{m_{t \bar t}^2}{m_t^2}, 
\ee
which, in the top-pair invariant mass, $m_{t\bar t}$, range relevant for the LHC, is subdominant to new physics effects as long as $\abs{\mu_t} \leq 0.1$. Therefore, the QCD contribution to the dipole operator can be neglected at first approximation, with the understanding that a full QCD NLO estimate, though computationally challenging, will be necessary if the limit is to be strengthened.

Finally, small sources of theoretical uncertainties come from the PDFs and the top-quark mass,
but these are negligible compared to the present level of precision. In fact, comparing the results obtained with two different PDF sets highlights a related uncertainty of the order of 0.1\%. This matches the magnitude of the uncertainty induced by the top-quark mass measurement, obtained by varying its mass within two standard deviations of its experimental value.
 
%%%%%%%%%%%%%%%%%%%%%%%%%%%%%%%%%%%%%%%%%%%%%%%%%%%%%%%%%
\subsection{Limits}

The experimental values of the correlation coefficients $\CC_{ij}$ and their uncertainties have been recently published by the CMS Collaboration in an analysis of data taken at the LHC at $\sqrt{s}=13$ TeV and an integrated luminosity of 138 fb$^{-1}$~\cite{CMS:2024zkc}. We focus on the events characterized by top-quark pairs with invariant mass $m_{\bar t t} > 800$ GeV and scattering angle $\Theta$ satisfying $|\cos \Theta| < 0.4$. These cuts define the highest bin in the energy with a scattering angle closest to $\pi/2$, for which we know~\cite{Aoude:2022imd,Fabbrichesi:2022ovb} that the effect of the new physics is the largest. Softening the angular cuts or extending the analysis to include different energy bins would then improve the result only marginally.

The tilded quantities appearing in \eq{pdf} are then integrated over the scattering angle and top-pair invariant mass to give analytic correlation coefficients that account for the extension of the experimental bin. The results obtained in the SM limit, $\mu_t=0$, are in good agreement with the central values extracted from the experimental data in the same bin. The uncertainties in the $\CC_{ij}$ coefficients are propagated to the chosen distance measure to give $\sigma_{\mathscr{D}^{T/F}}$ in \eq{chi2top}, thereby allowing for a test of the chromomagnetic moment of the top quark based on actual experimental data. 

\begin{figure} [h!]
    \centering
    \includegraphics[width=0.49\linewidth]{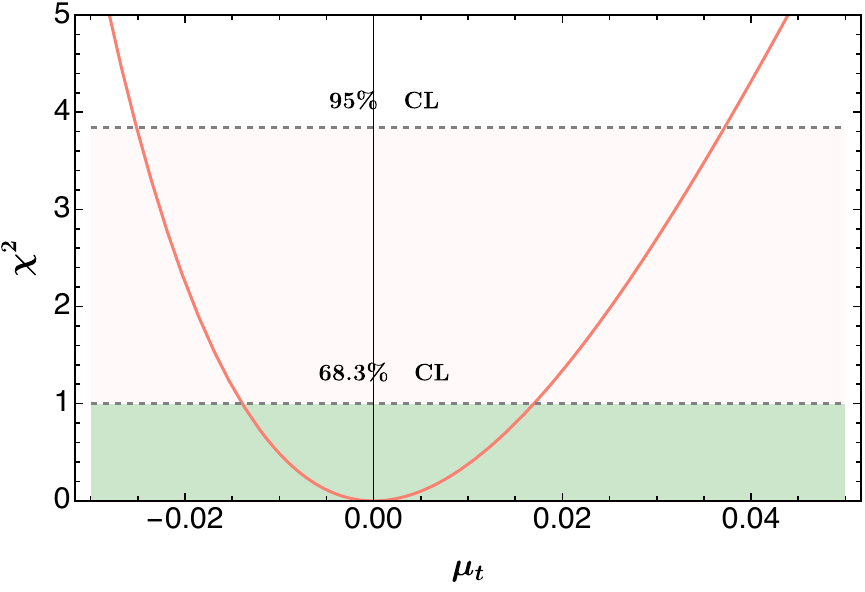}
    \includegraphics[width=0.49\linewidth]{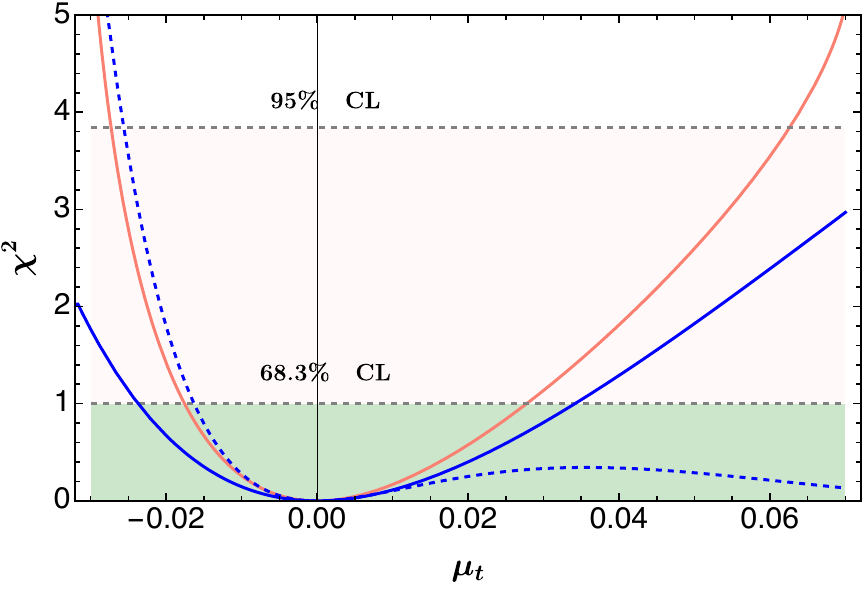}
    \caption{(left panel) $\chi^2$ test for the chromomagnetic dipole moment of the top quark $\mu_t$ obtained through the trace distance. For the sake of reference, in the right panel we show the constraining power of the fidelity distance (red line), concurrence (blue solid line, see \eq{concurrence}) and magic (blue dashed line, see \eq{eqm222}).}
    \label{fig:mutop}
\end{figure}

The 95\% (68\%) CL limits on the coefficient $\mu_t$ are then obtained by performing a $\chi^2$ test which quantifies, through the trace distance, the discrepancy between the SM density matrix and the density matrix that includes the contribution of the chromomagnetic dipole moment:
\be
\left( \frac{\mathscr{D}^{T/F}[\rho_{\text{\tiny NP}}(\mu_t), \rho_{\text{\tiny SM}}]}{\sigma_{\mathscr{D}^{T/F}}} \right)^2 \leq (1.00)\; 3.84 \,.  \label{chi2top}
\ee
In this first example we use the trace distance, or the fidelity distance, without the cross section because we want to gauge how it fares by itself when compared against the limits given by the cross section. Fig.~\ref{fig:mutop} shows the $\chi^2$ values obtained as the parameter $\mu_t$ varies over the considered range for the trace distance (left panel) and the fidelity distance (right panel, red line). As anticipated, the former provides the strictest bounds. For the sake of reference, in the right panel we also show the exclusion contours obtained by using the concurrence (solid blue line) and the magic (dashed blue lines), which we introduce in \eq{concurrence} and \eq{eqm222}, respectively. The corresponding 95\% confidence interval is reported in Table~\ref{tab:tops} together with a  determination based on the cross sections for top-quark pair production measured at LHC and Tevatron~\cite{Fabbrichesi:2013bca} that we take as our benchmark.

Though the relative uncertainties in the CMS measurement of the top-quark pair spin correlations are still rather large---as they are more than 10\%---the use of the trace distance already provides a bound on the chromomagnetic dipole moment comparable to that given by a combination of cross sections whose relative uncertainties are smaller---as they are of order 5\% \cite{Fabbrichesi:2013bca}. We expect that future measurements of the correlation coefficients will improve the precision of the experimental result and, accordingly, give even more stringent bounds on $\mu_t$. Regardless of these future improvements, the trace distance can already be a valuable addition if included into a global fit like that in Ref.~\cite{Brivio:2019ius}.

\begin{table}[t]
  \tablestyle[sansboldbw]
  \begin{tabular}{*{3}{p{0.3\textwidth}}}
  \theadstart
      \thead  Benchmark&\thead  This work&\thead  Best global fit \\
  \tbody
   $-0.046\leq\mu_{t} \leq 0.040 $ & $  -0.025 \leq|\mu_{t}| \leq 0.037$ & $-0.004\leq \mu_t \leq 0.007$ \\
  \hline%
  \tend
  \end{tabular}
  \caption{\footnotesize \label{tab:tops} \textrm{95\% CL limits for the chromomagnetic dipole moment of the top quark $\mu_t$. The benchmark value is taken from Ref.~\cite{Fabbrichesi:2013bca}, which uses the cross sections for top-quark pair production measured at LHC and Tevatron. The best current limit, $-0.004\leq \mu_t \leq 0.007$,  is obtained by simultaneously using cross sections of many different processes in a global fit~\cite{Brivio:2019ius}.}}
  \end{table}

%%%%%%%%%%%%%%%%%%%%%%%%%%%%%%%%%%%%%%%%%%%%%%%%%%%%%%%%%
\section{Anomalous couplings of the $\tau$ lepton at Belle}
%%%%%%%%%%%%%%%%%%%%%%%%%%%%%%%%%%%%%%%%%%%%%%%%%%%%%%%%%

{\versal The second example} we propose has to do with the $\tau$ lepton. The anomalous couplings of this particle are defined by means of the most general vertex, allowed by Lorentz and gauge symmetries, that couples the photon  (with  momentum $q$) to the $\tau$ lepton.  
The vertex, $\Gamma_\gamma^\mu$, can be written as
\be
-ie\,\bar \tau \,\Gamma^\mu(q^2) \,\tau \,A_\mu(q)
=
-ie \,\bar \tau \left[ \gamma^\mu F_1(q^2)  +  \frac{i \sigma^{\mu\nu}q_\nu}{2 m_\tau} F_2(q^2) 
+  \frac{\sigma^{\mu\nu} \gamma_5 q_\nu}{2 m_\tau} F_3(q^2)  \right]   \tau\, A_\mu(q) \, , \label{intGamma}
\ee
yielding the magnetic and electric dipole moments 
\be
a_\tau = F_2(0) \quad \mbox{and} \quad d_\tau = \frac{e}{2 m_\tau} F_3(0)\,,
\ee
as well as the mean squared radius of the $\tau$ lepton
\be
\langle \vec r^{\;2} \rangle = -  6 \left .\frac{d G_E}{d \vec q^{\;2}} \right|_{q^2=0} \,, \quad \text{with} \quad 
G_E(q^{2}) =  F_{1}(q^{2}) + \frac{q^{2}}{4 m_{\tau}^{2}} F_{2}(q^{2})\,. \label{tauradius}
\ee
In the following analysis we retain only the leading order terms in the expansions of the form factors $F_{2,3}(q^2)$ for $q^2\to 0$,  whereas for that of $F_1(q^2)$ we include also the first order correction in $q^2$.

Within the effective field theory framework, the form factors introduced in \eq{intGamma} yield $SU(2)_L \otimes U(1)_Y$ invariant effective operators involving the $\tau$ lepton. The largest anomalous contributions are governed by three higher-dimensional operators:
\be\label{O1}
{\cal  O}_1 = e \frac{C_1}{m_\tau^2} \bar \tau \gamma^\mu  \tau D^\nu F_{\mu\nu} \, ,\quad  {\cal O}_2= e \frac{C_2\,\upsilon}{2 m_\tau^2 } \bar \tau \sigma^{\mu\nu} \tau F _{\mu\nu} \, , \quad \mbox{and} \quad {\cal O}_3= e \frac{C_3\,\upsilon}{2 m_\tau^2 } \bar \tau \sigma^{\mu\nu} \gamma_{5}\tau F _{\mu\nu}\, ,
\ee
where $D^\nu$ is the covariant derivative, $F_{\mu\nu}$ the electromagnetic field strength tensor, and $\upsilon=174$ GeV is the rescaled Higgs field vacuum expectation value.  Left-handed and right-handed chiral fields contribute equally to the value of the dimensionless Wilson coefficients $C_i$, which are assumed to be real. 

The operator ${\cal O}_1$ encapsulates the leading $q^2$ dependence of the form factor $F_1$, while ${\cal O}_{2,3}$ give the zeroth order terms in the $q^2\to0$ expansion of the form factors $F_{2}$ and $F_3$:
\be 
\label{f12a}
F_1(q^2)=1+C_1 \frac{q^2}{m_\tau^2} +\ldots\qquad\mbox{and}\qquad F_{2,3}(0)=2\, C_{2,3} \frac{\upsilon}{m_\tau} \,.
\ee

%%%%%%%%%%%%%%
\begin{figure}[h!]
\begin{center}
\includegraphics[width=4.5in]{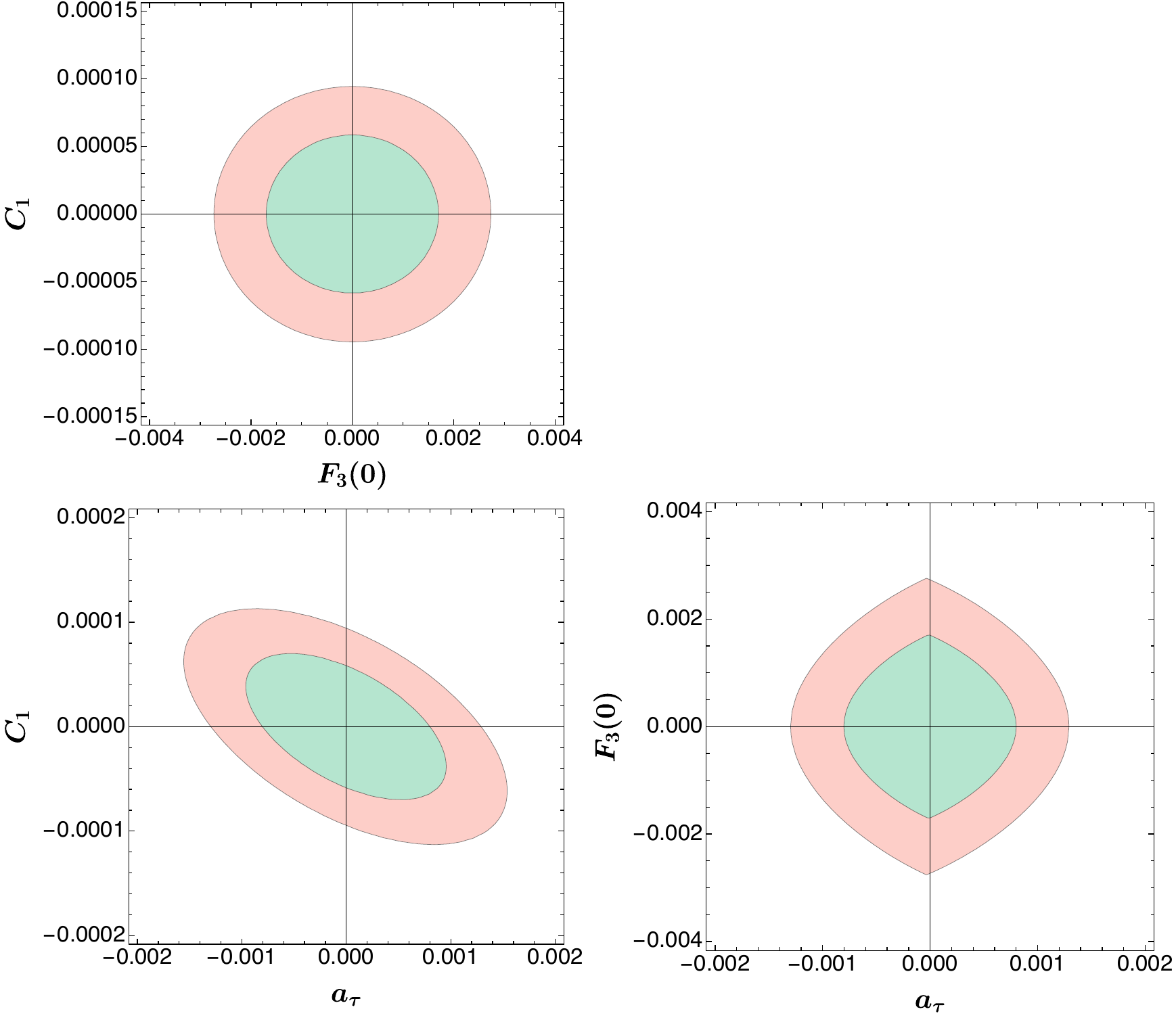}
\caption{\footnotesize 
\label{fig:limits_belle} Limits on the electromagnetic anomalous couplings of the $\tau$ lepton at Belle obtained with the test in \eq{chi2Belle}. 
}
\end{center}
\end{figure}
%%%%%%%%%%%%%%%%%%%%%%

The operators ${\cal O}_{2,3}$ include for the sake of convenience an extra factor of $\upsilon/m_\tau$, resulting from the dimension-six $SU(2)_L\times U(1)_Y$ gauge invariant operators involving the Higgs field after the electroweak symmetry breaking. Other operators of higher dimensions can in general contribute to the expressions of the form factors yielding  corrections of higher $q^2$ order that, in the considered limit, are negligible. 

The photon mediating the process  $e^+ e^- \to \tau^+\tau^-$ at Belle  is not on shell and the definitions above only apply if the difference between the form factors at $q^2\neq 0$ and their values at $q^2=0$ is sufficiently small. This is expected to be the case unless there is a threshold nearby.

\subsection{Limits}

The present experimental limit on the magnetic moment of the $\tau$ lepton is 
\be
-0.052  < a_\tau  <  0.013    \quad  \mbox{(95\%) CL} \; \text{\cite{DELPHI:2003nah}}\,,
\ee
which remains one order of magnitude above the SM prediction for this parameter
\be
a_\tau^{\rm SM}=1.17721(5) \times 10^{-3}\,\text{\cite{Eidelman:2007sb}} \label{aSM}.
\ee
The best current bounds on the electric dipole moment come instead from the Belle experiment, finding
\be
-0.185  \times 10^{-16}     < d_\tau < 0.061 \times 10^{-16} \; \mbox{e cm} \quad \mbox{(95\%) CL}\; \text{\cite{Belle:2021ybo}}
\ee
with an integrated  luminosity of 833 fb$^{-1}$ under the assumption that the parameter is real. Non-vanishing values would imply the presence of $CP$ violation.

The possibility of improving these limits has been discussed in the literature~\cite{PhysRevD.106.093007,Bernabeu:2008ii,BERNABEU2008160,Chen:2018cxt,Bernreuther:2021elu}, in particular a recent proposal used a combination of the entanglement, the cross section, and a $CP$-odd triple product of momenta and polarizations to constrain the parameters~\cite{Fabbrichesi:2024xtq}.  

\begin{table}[h!]
  \centering
  \tablestyle[sansboldbw]
  \begin{tabular}{p{0.4\linewidth} p{0.4\linewidth}}
  \theadstart
      \thead PDG (2022) &\thead  This work \\
  \tbody
 $-1.9 \times 10^{-17} \leq d_{\tau} \leq 6.1 \times 10^{-18} $ e cm& $  |d_{\tau}| \leq 1.5 \times 10^{-17} $ e cm \\
 $ -5.2 \times 10^{-2} \leq a_{\tau} \leq 1.3 \times 10^{-2}$ &  $ | a_{\tau} |\leq 1.0 \times 10^{-3} $   \\
 $\qquad \qquad\qquad |C_1| \leq 1.0 \times 10^{-5}$ & $ |C_{1}| \leq 1.0 \times 10^{-4}$ \\
  \hline%
  \tend
  \end{tabular}
  \caption{\footnotesize \label{tab:Belle} \textrm{Current experimental limits for the anomalous parameters in \eq{intGamma} ~\cite{Workman:2022ynf} and corresponding bounds obtained after marginalization of the 95\% joint confidence intervals shown in Fig.~\ref{fig:limits_belle}. The values in the second column are derived for a luminosity of 1 ab$^{-1}$ in the setup of the Belle experiment.}}
  \end{table}

In the present analysis we probe the anomalous couplings in \eq{intGamma} by using a $\chi^2$ test involving the trace distance between the SM density matrix and that inclusive of the anomalous couplings $C_1$, $ d_\tau$, and $a_\tau$. The cross section is included as well for the reason explained in Sec.~\ref{sec:distances}. We set the  $\chi^2$ test to indicate the joint (68\%) 95\% CL for two degrees of freedom as we vary the parameters in pairs away from the vanishing SM values:
\be
\left( \frac{\mathscr{D}^T[\rho_{\text{\tiny NP}}(a_\tau,\, d_\tau,\, C_1 ), \rho_{\text{\tiny SM}}]}{\sigma_{\mathscr{D}^T}} \right)^2+ \left[ 
\frac{\sigma_{\text{\tiny NP}} (a_\tau,\, d_\tau,\, C_1 )-\sigma_{\text{\tiny SM}}}{\sigma_{\sigma}}\right]^2 \leq (2.33)\; 5.99 \,.  \label{chi2Belle}
\ee
The value of the trace distance is obtained by computing the density matrix encoding the $\tau$-pair spin correlation for the SM and in the presence of the anomalous couplings. In the calculation we retain only the diagram mediated by a photon, as justified for a center of mass (CM) energy of $\sqrt s = 10.58$ GeV that mimics the Belle experiment setup. The uncertainties in the entries of the density matrix are taken from the Monte Carlo simulation in Ref.~\cite{Ehataht:2023zzt} and propagated into the uncertainty $\sigma_{\mathscr{D}^T}$ of the trace distance. The uncertainty $\sigma_{\sigma}$ in the cross section is obtained by rescaling the relative uncertainty of 0.3\%
on the integrated luminosity quoted in Ref.~\cite{Banerjee:2007is} for 833 fb$^{-1}$. All these uncertainties are at the $1\sigma$
CL and contain (to different extents) also systematic errors.

The results of the $\chi^2$ tests are shown in Fig.~\ref{fig:limits_belle}; the limits on the single parameters obtained via marginalization of the joint confidence intervals are quoted in the second column of Table~\ref{tab:Belle}. The PDG value for $C_1$ reported in the same table is derived from that on the contact interaction scale $\Lambda_{\text{C.I.}}$~\cite{Workman:2022ynf} via the following relation:
\be
C_1= \frac{6 \pi }{e^2}\, \qty(\frac{m_\tau}{ \Lambda_{\text{C.I.}}})^2\, . 
\label{eq:lambda}
\ee
In comparing the results for the coefficient $C_1$, we must bear in mind that $\Lambda_{\text{C.I.}}$ in \eq{eq:lambda} is taken at energies of about 200 GeV, whereas in our case the scale is effectively one order of magnitude smaller---about 10 GeV. Depending on the origin of the relevant contact interaction, the scaling of the related operator is at least linear in the energy, implying that our result should be compared with the approximated rescaled value $|C_1| \leq 1.0 \times 10^{-3}$ when assessing the power of the method.  

%%%%%%%%%%%%%%%%%%%%%%%%%%%%%%%%%%%%%%%%%%%%%%%%%%%%%%%%%
\subsection{Spin formalism and quantum tomography}

The experimental reconstruction of the spin density matrix is typically performed via the determination of the involved Fano coefficients. These are given by the averages of angular distributions of final state particles that reveal the orientation of the spin vectors of interest. The obtained averages, necessarily weighted by the angular distributions of the collected events, are taken over the ranges of the kinematic variables defined by the binning used in the reconstruction. Measuring, for instance, the spin state characterizing an ensemble of top-pairs emitted with a scattering angle close to $\Theta=\pi/2$---where new physics effects and quantum observables are often maximized---then requires the implementation of an angular cut which lowers the number of events used in the analysis and, consequently, its precision.  As we show below, this problem can be circumvented by using the spin formalism of helicity amplitudes.      

The theoretical framework of the spin formalism~\cite{Jacob:1959at,Leader:2011vwq} provides an alternative way to perform quantum tomography that leverages the properties of helicity states. The method decomposes the underlying production amplitudes into partial waves each characterized by a specific dependence on the scattering angle in the CM frame. The full reconstruction of the spin density matrix is then given in terms of the helicity amplitudes of the same production process. As an explicit example, the density matrix describing the spin state of $\tau$-lepton pairs created in the annihilation of unpolarized electron and positron beams via a photon exchange is given, in this formalism, by    

\be
\small
\rho \propto 
\begin{pmatrix}
 w_{-\frac{1}{2}\,\frac{1}{2}}\,w_{-\frac{1}{2}\frac{1}{2}}^* \,f_\Theta 
 & 
 w_{-\frac{1}{2}\,\frac{1}{2}}\,w_{-\frac{1}{2}\,-\frac{1}{2}}^*\,\frac{s_\Theta c_\Theta}{\sqrt{2}} 
 & 
 w_{-\frac{1}{2}\,\frac{1}{2}}\,w_{\frac{1}{2}\,\frac{1}{2}}^*\,\frac{s_\Theta c_\Theta}{\sqrt{2}} 
 & 
 w_{-\frac{1}{2}\,\frac{1}{2}}\,w_{\frac{1}{2}\,-\frac{1}{2}}^* \,\frac{s_{\Theta}^2}{2}  
 \\
 w_{-\frac{1}{2}\,-\frac{1}{2}}\,w_{-\frac{1}{2}\,\frac{1}{2}}^*\,\frac{s_\Theta c_\Theta }{\sqrt{2}} 
 & 
 w_{-\frac{1}{2}\,-\frac{1}{2}} \,w_{-\frac{1}{2},-\frac{1}{2}}^*\,s_{\Theta}^2 
 & 
 w_{-\frac{1}{2}\,-\frac{1}{2}} \,w_{\frac{1}{2}\,\frac{1}{2}}^*\, s_{\Theta}^2 
 &
 -w_{-\frac{1}{2}\,-\frac{1}{2}}\,w_{\frac{1}{2}\,-\frac{1}{2}}^*\,\frac{s_\Theta c_\Theta }{\sqrt{2}} 
 \\
 w_{\frac{1}{2}\,\frac{1}{2}}\,w_{-\frac{1}{2}\,\frac{1}{2}}^*\,\frac{s_\Theta c_\Theta}{\sqrt{2}} 
 & 
 w_{\frac{1}{2}\,\frac{1}{2}} \,w_{-\frac{1}{2}\,-\frac{1}{2}}^*\,s_{\Theta}^2 
 & 
 w_{\frac{1}{2}\,\frac{1}{2}} \, w_{\frac{1}{2}\,\frac{1}{2}}^*\,s_{\Theta}^2 
 & 
 -w_{\frac{1}{2}\,\frac{1}{2}}\,w_{\frac{1}{2}\,-\frac{1}{2}}^*\,\frac{s_\Theta c_\Theta }{\sqrt{2}} 
\\
w_{\frac{1}{2}\,-\frac{1}{2}}\,w_{-\frac{1}{2}\,\frac{1}{2}}^*\,\frac{s_{\Theta}^2}{2}  
& 
-w_{\frac{1}{2}\,-\frac{1}{2}}\,w_{-\frac{1}{2}\,-\frac{1}{2}}^*\,\frac{s_\Theta c_\Theta}{\sqrt{2}} 
& 
-w_{\frac{1}{2}\,-\frac{1}{2}}\,w_{\frac{1}{2}\,\frac{1}{2}}^*\,\frac{s_\Theta c_\Theta}{\sqrt{2}} 
& 
w_{\frac{1}{2}\,-\frac{1}{2}} \,w_{\frac{1}{2}\,-\frac{1}{2}}^*\,f_\Theta    \label{eq:rhoHA}
\end{pmatrix}\, ,
\ee
in which $f_\Theta = (3 + \cos 2 \Theta )/4$, $s_\Theta = \sin \Theta$, $c_\Theta = \cos \Theta$, and $\Theta$ is the scattering angle of the $\tau^+$ lepton in the CM frame. The omitted proportionality factor ensures that $\Tr(\rho)=1$. Each entry of the above density matrix is determined by a product of helicity amplitudes, resulting in a combination of the helicity coefficients $w_{ij}$, with $i,j=\pm1/2$, which multiplies the Wigner function set by the conservation of angular momentum~\cite{Jacob:1959at,Leader:2011vwq}. The unconventional helicity ordering reflects the fact that, rather than the usual helicity frames, we used a common frame of reference to decompose the spin vectors in the calculation. This choice facilitates the comparison with the results recently obtained in the literature, which use a unique $n,\,r,\,k,$ triad to decompose the spin vectors of the particles under examination.     
     
The spin state of the system under study can then be fully reconstructed at collider experiments by simply measuring the individual helicity amplitudes for the $e^+ e^-\to \tau^+ \tau^-$ process in a kinematical regime in which the $Z$ boson contribution is negligible. Once each entry of the density matrix is measured, the products of helicity coefficients appearing in \eq{eq:rhoHA} can be obtained by computing the average of the Wigner functions over the scattering angle values used in the observation, weighted by the angular distribution of the events. 

The advantage of this method consists in that the products of helicity coefficients can be experimentally determined by using all the collected events pertaining to the process of interest, without introducing angular cuts that would reduce the precision of the result. The known angular dependence of the density matrix entries then provides the means to obtain the operator describing the spin state of an ensemble of particles given for an arbitrary value, or range, of the scattering angle $\Theta$. 

%%%%%%%%%%%%%%
\begin{figure}[h!]
\begin{center}
\includegraphics[width=3.5in]{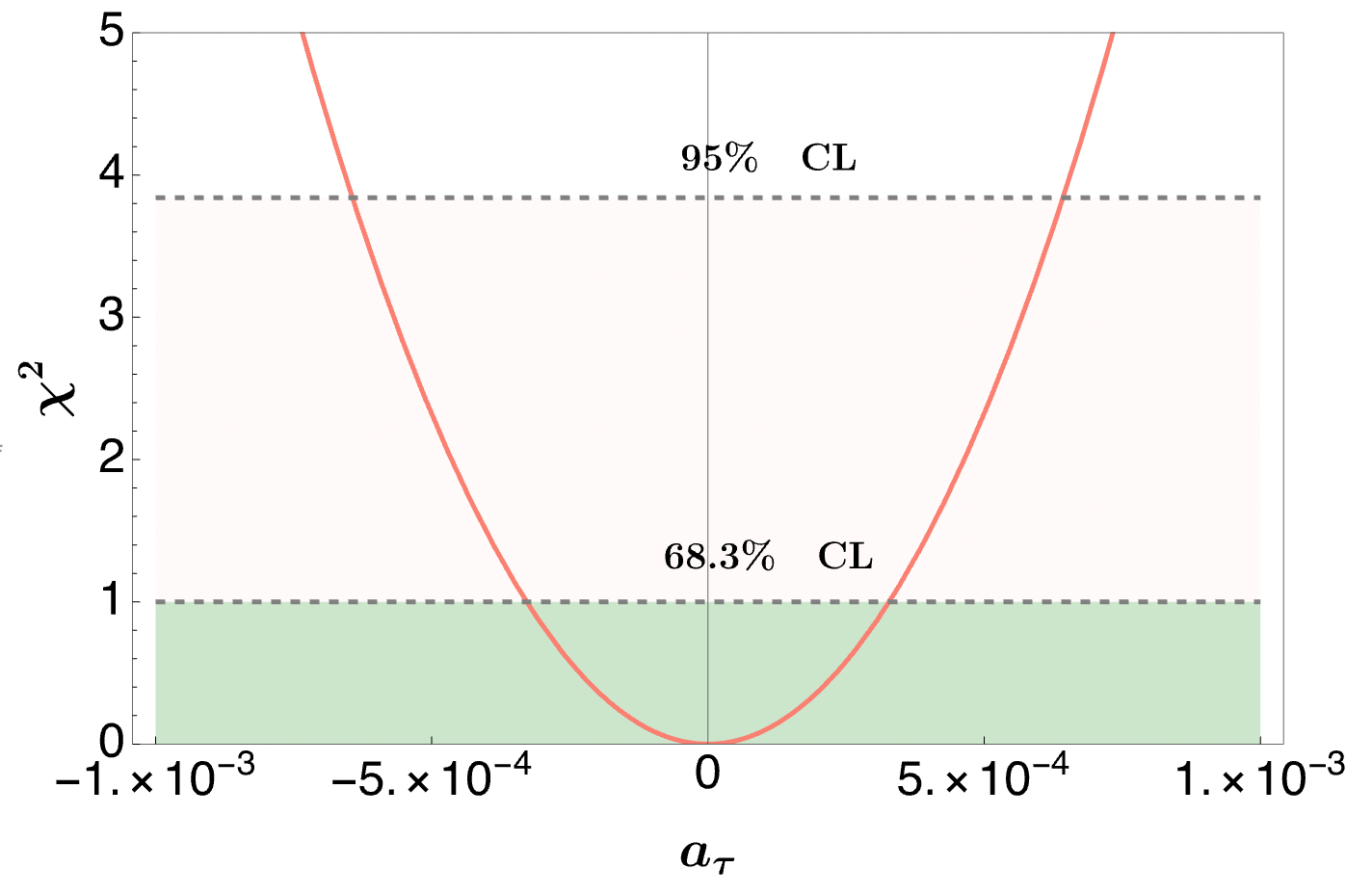}
\caption{\footnotesize 
\label{fig:HA} The $\chi^2$ confidence intervals obtained for the parameter $a_\tau$ by using the helicity amplitude method to compute the density matrix at a scattering angle $\Theta=\pi/2$.
}
\end{center}
\end{figure}
%%%%%%%%%%%%%%%%%%%%%%

An an example of the helicity amplitude method, we now constrain the anomalous magnetic dipole moment, $a_\tau$, of the $\tau$ pairs described by the density matrix in \eq{eq:rhoHA} for a scattering angle $\Theta=\pi/2$  that maximizes new physics effects. In place of the experimental results, we use the analytic expression involved helicity coefficients. In this case we find
 \be
 w_{\frac{1}{2}\, \frac{1}{2}}=w_{-\frac{1}{2}\, -\frac{1}{2}} \propto 4\,m_\tau^2 + a_\tau s \quad \text{ and} \quad 
  w_{\frac{1}{2}\, -\frac{1}{2}} = w_{-\frac{1}{2}\, \frac{1}{2}} \propto 2 \, i\, \sqrt{2 s} \, m_\tau \,(1 + a_\tau)\,,\label{eq:HAinput}
  \ee
omitting an overall normalization factor that cancels out in \eq{eq:rhoHA} once the matrix is normalized.

Having the helicity coefficients, we then compute the density matrix for $\Theta=\pi/2$ and repeat the analysis in the previous section. The statistical uncertainties that enter our $\chi^2$ test are the same as before because the computation of the helicity coefficients uses the full range of the scattering angle; they are also the smallest achievable from angular distributions, barring systematic errors, as they are obtained with the largest number of  available events. 

The result of the $\chi^2$ test is shown in Fig.~\ref{fig:HA} while the limits obtained for $a_\tau$ are reported in Table~\ref{tab:Belle2}. The improvement with respect to the result in Table~\ref{tab:Belle}  recovers the bound given in Ref.~\cite{Fabbrichesi:2024xtq} obtained by imposing the cut $|\cos\Theta|\leq0.4$ on the scattering angle. Given the large number of events that pass the cut used in the setup of Ref.~\cite{Fabbrichesi:2024xtq},  the spin formalism cannot dramatically improve the reach of the analysis. We nevertheless expect that the use of the  helicity amplitude method could make a real difference in analyses with low numbers of event in the scattering angle bins of interest.

We note that the application of the spin formalism bears some similarity to the kinematic method of Ref.~\cite{Cheng:2024rxi}.  With this method one parametrizes the density matrix, or equivalently the spin correlation matrix and the polarization vectors, by its dependence on the scattering angle and the scattering energy.  This analytic dependence is used to reconstruct the density matrix directly from kinematics, without the use of angular distributions.  In contrast, the spin formalism starts with reconstruction using angular distributions, but then uses the analytic dependence of the density matrix to convert the measurement of the quantum state in one phase space region to the measurement of the quantum state of a different phase space region.

\begin{table}[t]
  \tablestyle[sansboldbw]
  \begin{tabular}{*{2}{p{0.35\textwidth}}}
  \theadstart
      \thead  Benchmark&\thead  This work, II \\
  \tbody
   $-5.2 \times 10^{-3}\leq a_{\tau} \leq 1.3 \times 10^{-3} $ & $  |a_{\tau}| \leq 6.4 \times 10^{-4}$ \\
  \hline%
  \tend
  \end{tabular}
  \caption{\footnotesize \label{tab:Belle2} \textrm{95\% CL limits for the magnetic dipole moment of the $\tau$ lepton $a_\tau$ obtained by the helicity amplitude method. }}
  \end{table}

%%%%%%%%%%%%%%%%%%%%%%%%%%%%%%%%%%%%%%%%%%%%%%%%%%%%%%%%%
\section{Anomalous couplings of the $\tau$ lepton at LEP3}
%%%%%%%%%%%%%%%%%%%%%%%%%%%%%%%%%%%%%%%%%%%%%%%%%%%%%%%%%

{\versal The last example} we propose considers again  the $\tau$ lepton and its anomalous couplings, albeit in the setup of Ref.~\cite{Fabbrichesi:2024wcd} which sets the CM energy at the $Z$ boson mass.

In order to investigate the anomalous couplings we write again the most general electroweak Lorentz-invariant vertex involving, this time, the $Z$ boson and the $\tau$ lepton. Barring operators of dimension higher than five, the vertex $\Gamma_Z^\mu$ can be written as
\begin{multline}
i\,\frac{g}{2\,\cos \theta_W} \,\bar \tau \,\Gamma^\mu(q^2) \,\tau \,Z_\mu(q)
=\\
i \,\,\frac{g}{2\,\cos \theta_W} \, \bar \tau \left[ \gamma^\mu F_1^V(q^2)  +\gamma^\mu \gamma_5 F_1^A(q^2) +\frac{i \sigma^{\mu\nu}q_\nu}{2 m_\tau} F_2(q^2) 
+  \frac{\sigma^{\mu\nu} \gamma_5 q_\nu}{2 m_\tau} F_3(q^2)  \right]   \tau\, Z_\mu(q) \, , \label{intZ}
\end{multline}
with $F_1^V(0) = g_V=-1/2 + 2\, \sin^2\theta_W$, $F_1^A(0) = -g_A=1/2$, and with $F_2(0)$ and $F_3(0)$ being the magnetic and the electric dipole moments, respectively. We parametrize the expansions in $q^2\to 0$ of the first two form factors as
\be
F_1^{V,A} (q^2) = F_1^{V,A}(0) + \frac{q^2}{m_Z^2} C_1^{V,A}\, ,
\ee
and give limits on the coefficients $ C_1^{V,A}$ at $q^2=m_Z^2$, which induce the anomalous couplings.

%%%%%%%%%%%%%%%%%%%%%%%%%%%%%%%%%%%%%%%%%%%%%%%%%%%%%%%%%
\begin{figure}[h!]
    \centering
    \includegraphics[width=6in]{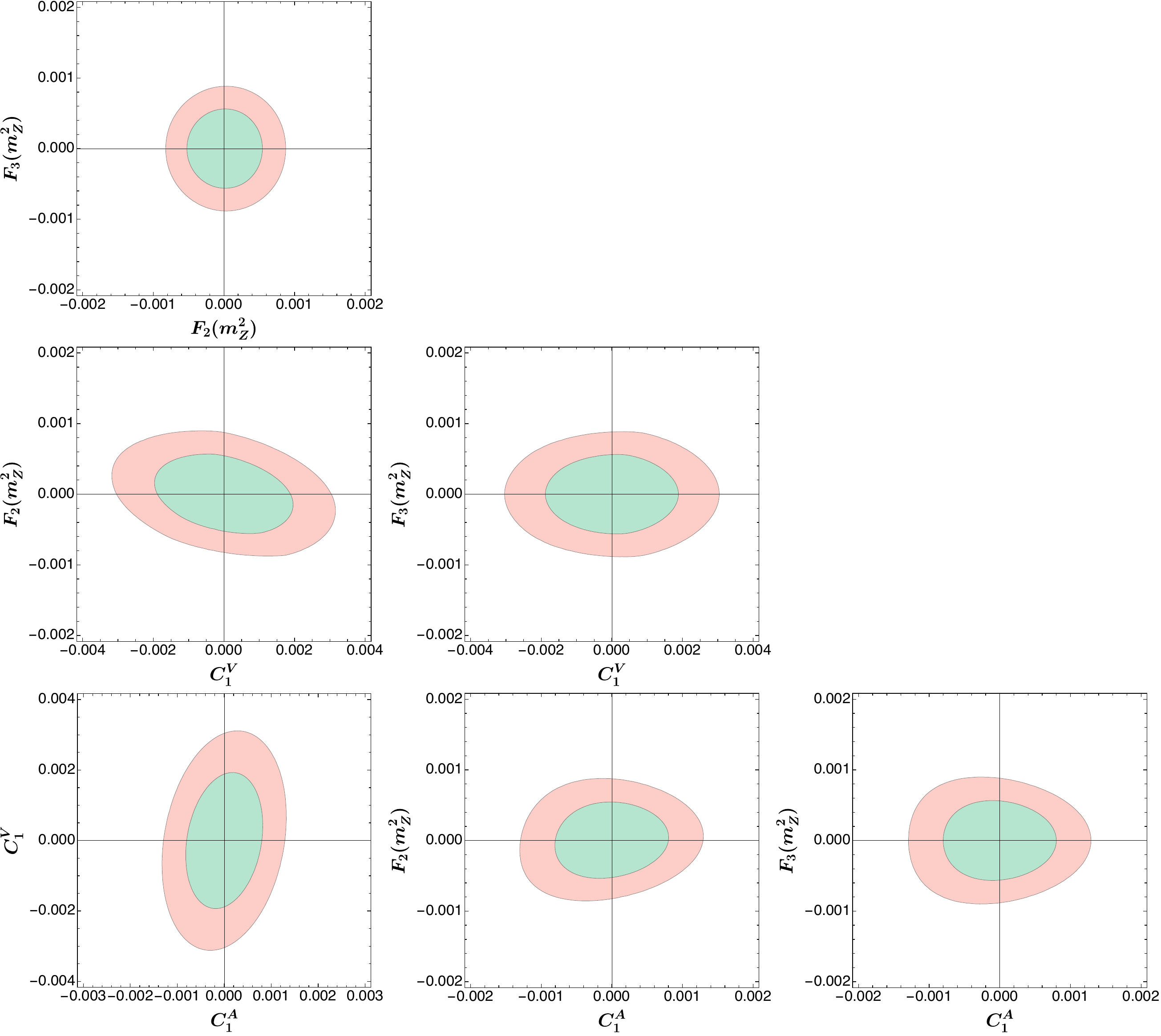}
    \caption{Limits on the four form factors entering the coupling of the $\tau$ lepton to the $Z$ boson obtained from the $e^+ e^-\to Z\to \tau^+ \tau^-$ process at the Z peak via quantum tomography. The shown joint confidence intervals use the test in \eq{chi2LEP3}.}
    \label{fig:LEP3}
\end{figure}
%%%%%%%%%%%%%%%%%%%%%%%%%%%%%%%%%%%%%%%%%%%%%%%%%%%%%%

As before, the form factors introduced in \eq{intZ} yield $SU(2)_L \times U(1)_Y$ invariant effective operators for the $\tau$ lepton in the language of effective field theory. The leading contributions are given by four higher-dimensional operators, written here in the Warsaw basis~\cite{Grzadkowski:2010es}:
\bea\label{O2}
{\cal O}_1 &=& e \frac{C_1^V}{m_\tau^2} \bar \tau \gamma^\mu  \tau D^\nu Z_{\mu\nu} \, ,\quad  
{\cal O}_1^\prime = e \frac{C_1^A}{m_\tau^2} \bar \tau \gamma^\mu \gamma^5  \tau D^\nu Z_{\mu\nu} \, , \\
{\cal O}_2&=& e \frac{C_2\,\upsilon}{2 m_\tau^2 } \bar \tau \sigma^{\mu\nu} \tau Z_{\mu\nu} \, ,\quad \mbox{and} \quad {\cal  O}_3= e \frac{C_3\,\upsilon}{2 m_\tau^2 } \bar \tau \sigma^{\mu\nu} \gamma_{5}\tau Z_{\mu\nu}\, ,
\eea
where $D^\nu$ is the covariant derivative, $Z_{\mu\nu}$ the field strength tensor of the $Z$ field and $\upsilon=v/\sqrt{2}=174$ GeV the rescaled Higgs field vacuum expectation value. The dimensionless Wilson coefficients $C_i$ in \eq{O2} are assumed to be real. 

The operators ${\cal O}_1$ and ${\cal O}_1^\prime$ give the leading $q^2$ dependence of the form factors $F_1^{V,A}$, while ${\cal O}_{2,3}$ produce the $q^2=m_Z^2$ terms in the $q^2$ expansion of the $F_{2,3}$ form factors:
\be 
\label{f12b}
F_1^{V,A}(q^2)=1+C^{V,E}_1 \frac{q^2}{m_\tau^2} +\ldots\qquad\mbox{and}\qquad F_{2,3}(m_Z^2)=2\, C_{2,3} \frac{\upsilon}{m_\tau} \,.
\ee
The operators ${\cal O}_{2,3}$ include again the extra factor $\upsilon/m_\tau$, sourced by the dimension-six operators involving the Higgs field after the electroweak symmetry breaking. Operators of higher dimensions yield higher order terms in the expansion of the form factors that are negligible in the present context. 

LEP3 is a possible future $e^+ e^-$ collider that could be built inside the existing LHC tunnel by installing new electromagnetic cavities and magnets~\cite{Blondel:2012ey}, thereby achieving a CM energy of about $\sqrt{s}=240$ GeV and a luminosity of $10^{34}$ cm$^{-2}$ s$^{-1}$. It would make it possible to study in detail the Higgs boson couplings and other processes at the energy of the $Z$ boson mass by copiously producing the gauge boson.

%%%%%%%%%%%%%%%%%%%%%%%%%%%%%%%%%%%%%%%%%%%%%%%%%%%%%%%%%
\subsection{Limits}

%%%%%%%%%%%%%%%%%%%%%  
   \begin{table}[h!]
  \tablestyle[sansboldbw]
  \begin{tabular}{p{0.35\linewidth} p{0.35\linewidth}}
  \theadstart
       \thead  \hspace{0.2cm}  Benchmark   \hspace{0.2cm}   \hspace{0.2cm} &      \thead    \hspace{0.5cm} This work   \hspace{0.2cm} \tabularnewline
  \tbody
$  -0.002 \leq F_2(m_Z^2) \leq  0.003 $ &  $ \qquad |F_2(m_Z^2)| \leq  0.001$\\
$ \qquad \qquad  |F_3(m_Z^2)| \leq 0.001$ & $ \qquad |F_3(m_Z^2)| \leq 0.001$ \\
$ -0.009 \leq C_1^V \leq 0.010$  &$ \qquad |C_1^V| \leq 0.003$  \\
$\qquad \qquad |C_1^A| \leq 0.001 $ & $ \qquad | C_1^A| \leq 0.001 $\\
  \tend
  \end{tabular}
  \caption{\footnotesize \label{tab:LEP3} \textrm{Bounds obtained at the 95\% CL for the form factors through the marginalization of the joint confidence intervals shown in Fig.~\ref{fig:LEP3}. The benchmark values are taken from Ref.~\cite{Fabbrichesi:2024wcd}. }}
  \end{table}
 %%%%%%%%%%%%%%%%%%%

Currently there are no experimental limits on the form factors in \eq{intZ}. A recent estimate~\cite{Fabbrichesi:2024wcd} uses a combination of the entanglement, the total cross section and a triple product correlation to give an estimate that we use as benchmark to gauge the improvement obtained by using the trace distance.

\begin{figure}[h!]
    \centering
    \includegraphics[width=6in]{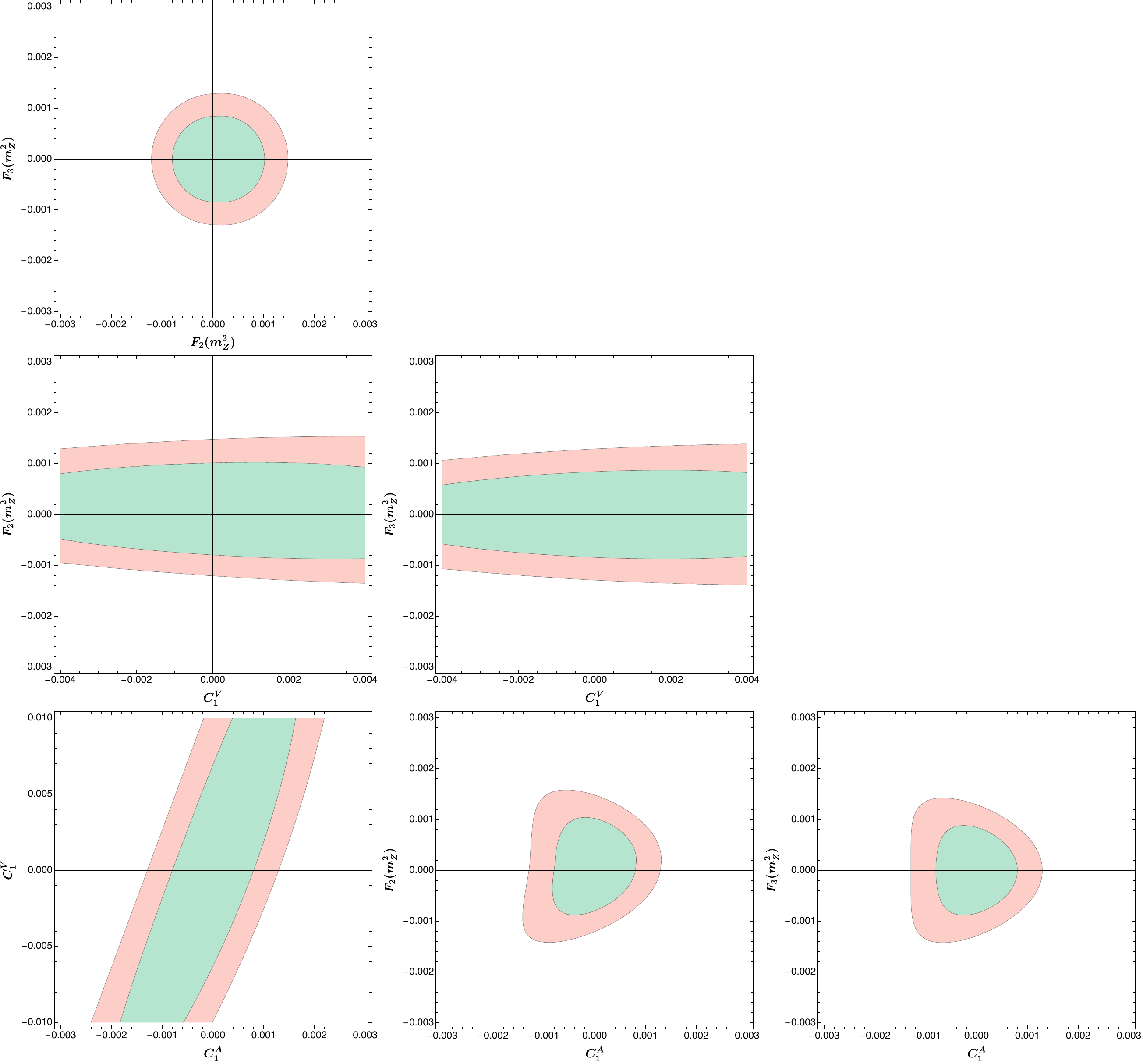}
    \caption{Limits on the four form factors entering the coupling of the $\tau$ lepton to the $Z$ boson obtained from the $e^+ e^-\to Z\to \tau^+ \tau^-$ process at the Z peak via quantum tomography. The shown joint confidence intervals use the test in \eq{chi2LEP3}, albeit the replacement of the trace distance with the fidelity distance in \eq{DF}.}
    \label{fig:LEP3FD}
\end{figure}

For the analysis we compute analytically both the density matrix for the SM and that inclusive of the new physics effects, starting with the amplitudes of the $e^+ e^-\to\gamma,\, Z\to \tau^+ \tau^-$ process, in which we retain both the intermediate contributions and set $\sqrt s = m_Z$ with $m_Z$ being the $Z$ boson mass. The results are then used in the  $\chi^2$ test: 
\begin{multline}
\left( \frac{\mathscr{D}^T[\rho_{\text{\tiny NP}}(C_1^V,\, C_1^A,\, F_2(m_Z^2),\, F_3(m_Z^2) ), \rho_{\text{\tiny SM}}]}{\sigma_{\mathscr{D}^T}} \right)^2 \\ + \left[ 
\frac{\sigma_{\text{\tiny NP}} (C_1^V,\, C_1^A,\, F_2(m_Z^2),\, F_3(m_Z^2)) -\sigma_{\text{\tiny SM}}}{\sigma_{\sigma}}\right]^2 \leq (2.33)\; 5.99 \,,  \label{chi2LEP3}
\end{multline}
with the uncertainties affecting the Fano coefficients, yielding $\sigma_{\mathscr{D}^T}$, and the cross section set according to the results of the Monte Carlo simulation in Ref.~\cite{Fabbrichesi:2024wcd}. Fig.~\ref{fig:LEP3} shows the limits obtained at a (68\%) 95\% CL with the test in \eq{chi2LEP3} by varying at a single time two of the new physics parameters $C_1^V$, $C_1^A$, $F_2(m_Z^2)$, and $F_3(m_Z^2)$. The bounds reported in Table~\ref{tab:LEP3} for the each parameter are obtained through marginalization. For the sake of comparison, we present in Fig.~\ref{fig:LEP3FD} the corresponding exclusion plots obtained by using the fidelity distance, \eq{DF}, in place of the trace distance in the test of \eq{chi2LEP3}.

Two of the limits found in this work match the corresponding benchmark values: those for $F_3(m_Z^2)$ and the ones for $C_1^A$. For the first parameter, Ref.~\cite{Fabbrichesi:2024wcd} used $CP$-odd triple products of momenta and polarization vectors that directly test the Fano coefficients of the involved density matrices, in combination, mimicking the power of the trace distance. The bounds on $C_1^A$ come instead from the total production cross section used both in the benchmark analysis and in this work. The limits affecting the remaining two form factors, $F_2(m_Z^2)$ and $C_1^V$, see a 30\% improvement obtained with the proposed methodology.

%%%%%%%%%%%%%%%%%%%%%%%%%%%%%%%%%%%%%%%%%%%%%%%%%%%%%%%%%
%%%%%%%%%%%%%%%%%%%%%%%%%%%%%%%%%%%%%%%%%%%%%%%%%%%%%%

%%%%%%%%%%%%%%%%%%%%%%%%%%%%%%%%%%%%%%%%%%%%%%%%%%%%%%%%%
\section{Quantum information observables and distances}
%%%%%%%%%%%%%%%%%%%%%%%%%%%%%%%%%%%%%%%%%%%%%%%%%%%%%%%%%

{\versal We conclude the paper} by briefly reviewing the quantum information observables used in the literature to constrain new physics effects and by gauging their power against the proposed distance measures. 

The concurrence, $\mathscr{C}$, quantifies the amount of entanglement present in a quantum state. The parameter ranges on the interval $0\leq \mathscr{C} \leq 1$.  A vanishing value signals the presence of a separable state---the absence of entanglement---while the upper bound is saturated by maximally entangled states such as Bell states. 

\begin{figure}[h!]
\begin{center}
\includegraphics[width=3.2in]{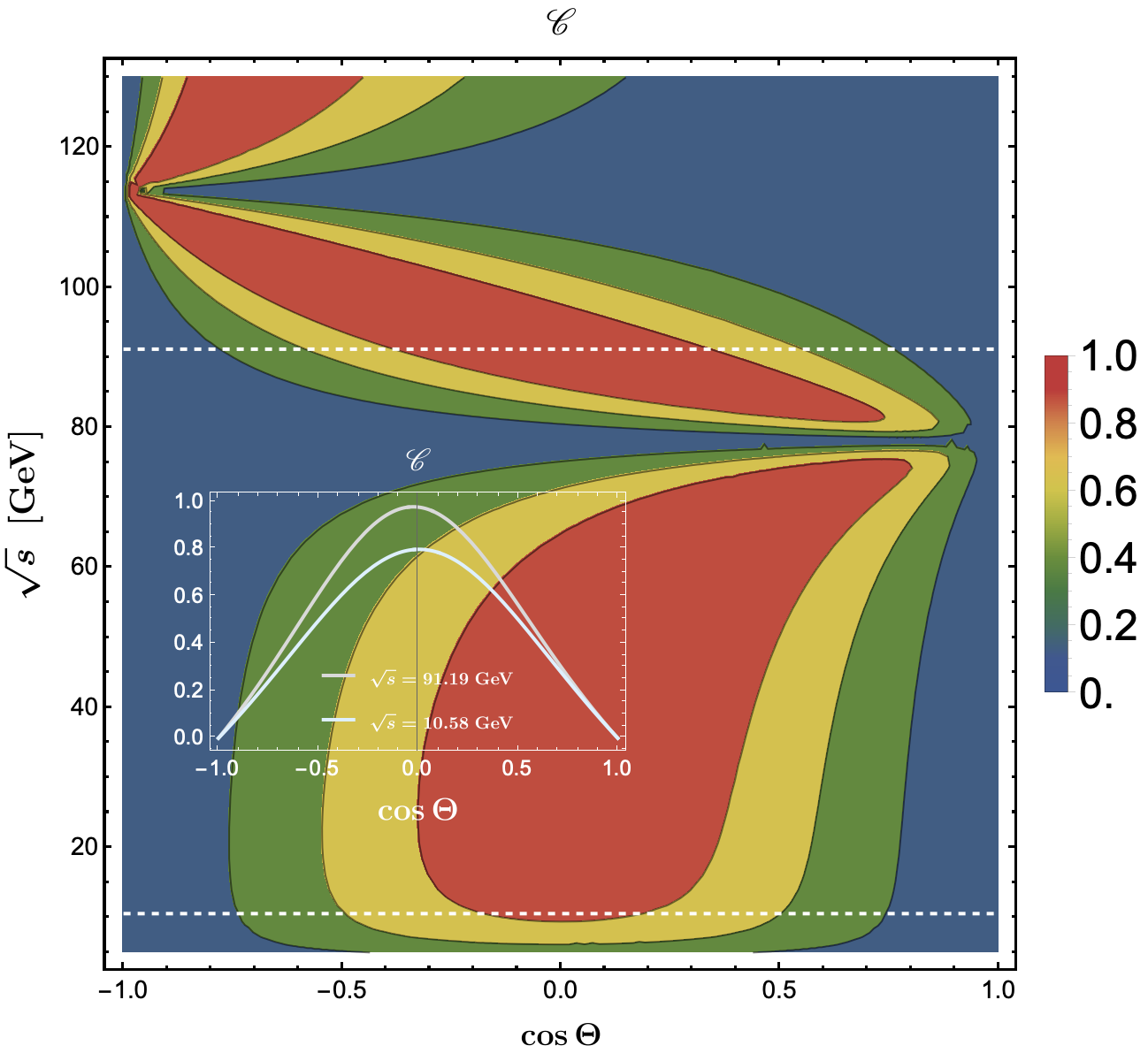}
\hspace{0.25cm}
\includegraphics[width=3.2in]{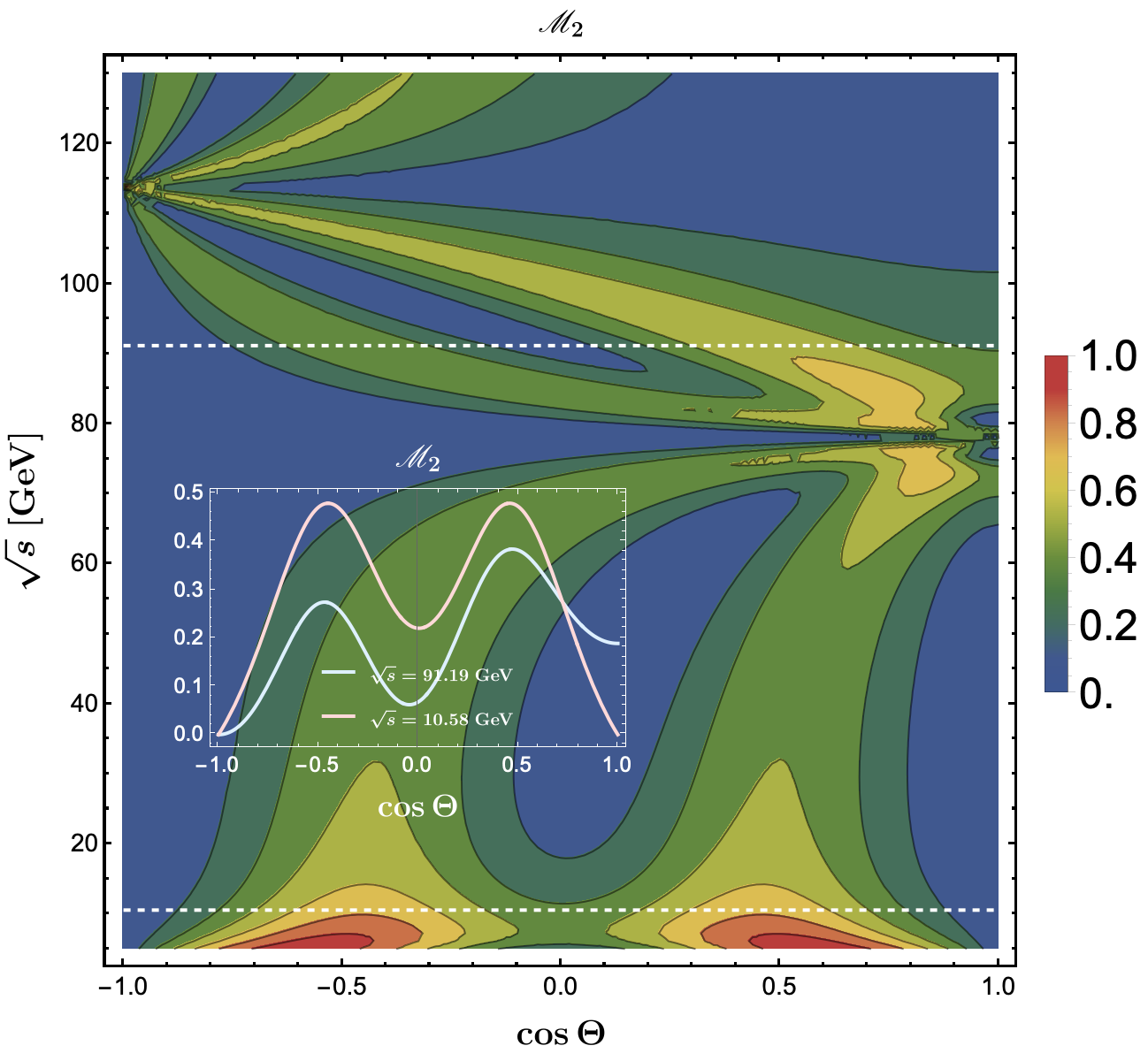}

\caption{\footnotesize 
\label{fig:kine} Concurrence and magic profiles over the entire kinematic space spanned by $\sqrt{s}$ and $\cos \Theta$ for the process $e^+ e^-\to\tau^+\tau^-$. The dashed white horizontal lines mark the center of mass energies of Belle (lower line) and LEP3 (upper line). The insets show the dependence on the scattering angle at these two energies.
}
\end{center}
\end{figure}

In the case of bipartite system formed by two qubits, such as a pair of spin-1/2 fermions, the concurrence can be analytically computed through the auxiliary matrix
\begin{equation} 
R=\rho \,  (\sigma_y \otimes \sigma_y) \, \rho^* \, (\sigma_y \otimes \sigma_y)\, ,
\label{auxiliary-R}
\end{equation}
where $\rho^*$ is the matrix obtained from a density matrix $\rho$ via the complex conjugation of the entries. Although non-Hermitian, the matrix $R$ possesses non-negative eigenvalues. By denoting with $r_i$, $i=1,2,3,4$, the square roots of the eigenvalues and assuming $r_1$ to be the largest, the concurrence of the state $\rho$ can be expressed as~\cite{Wootters:PhysRevLett.80.2245}
\begin{equation}
\mathscr{C}[\rho] = \max \big( 0, r_1-r_2-r_3-r_4 \big)\ .
\label{concurrence}
\end{equation}
The values of the concurrence of the final leptons for the $e^+e^-\to\tau^+\tau^-$ process, including both the $Z$ boson and the photon contributions, are given in the first panel of Fig.~\ref{fig:kine} over the kinematic space of the process.  

%%%%%%%%%%%%%%
%%%%%%%%%%%%%%%%%%%%%%

Magic is a resource that quantifies the importance of a given quantum state for quantum computation purposes. A quantum computation performed by using strongly entangled states does not give quantum computers an advantage over their classical counterparts. States characterized by a large value of magic, instead, allow these quantum machines to outperform ordinary computers~\cite{Aaronson_2004}.

Magic has been proposed as a useful observable at colliders~\cite{White:2024nuc}.  The parameter is usually defined in terms of the Stabiliser R\'enyi Entropies $\mathscr{M}_q$, a set of functions 
\be
\mathscr{M}_q = \frac{1}{1-q} \, \log_2 (\zeta_q)\, ,\quad \text{with} \quad \zeta_q = \sum_{P\in\mathcal{P}} \frac{\langle \psi| P | \psi \rangle^{2 q}}{2^n} \, ,
\ee
spanned by the integer value $q\geq2$. The quantity $\zeta_q$ is the weighted sum of the expectation values $\expval{P}{\psi}$  over the set $\mathcal{P}$ of Pauli strings $P$ needed for the stabilization of the state $\ket{\psi}$. In terms of these functions, magic is given by the second Stabiliser R\'enyi Entropy $\mathscr{M}_2$. 

For a bipartite qubit system described by a density matrix $\rho$, the above definition reduces to 
\begin{equation}
\mathscr{M}_2(\rho) = - \log_2 \qty(\frac{\sum_{P\in\mathcal{P}} \Tr^4(\rho P)}{\sum_{P\in\mathcal{P}} \Tr^2(\rho P)})\,,
\end{equation}
with $\mathcal{P} =\{\sigma_i\otimes\sigma_j\}$ and $i,j = 0,1,2,3$, having defined $\sigma_0$ as the identity matrix of dimension two. In terms of the Fano coefficients in \eq{eq:rho_deco} we then obtain
\begin{equation}
\mathscr{M}_2(\rho) = - \log_2 \qty[\frac{1+\sum_{i=1}^3\qty(\BB_i^+)^4+\sum_{i=1}^3\qty(\BB_i^-)^4+\sum_{i,j=1}^3\qty(\CC_{ij})^4}{1+\sum_{i=1}^3\qty(\BB_i^+)^2+\sum_{i=1}^3\qty(\BB_i^-)^2+\sum_{i,j=1}^3\qty(\CC_{ij})^2}]\,.
\label{eqm222}
\end{equation}
 We show how magic, computed for the polarizations of the final leptons, varies in the kinematic space of the $\tau$-pair production in the second panel of Fig.~\ref{fig:kine}. The behavior is remarkably different form that of the concurrence, shown in the first panel, as the parameters assume their largest values in complementary regions of the  kinematic space. This behavior makes these two observables almost complementary to each other in regions where there are non-negligible quantum effects.

%%%%%%%%%%%%%%%%%%%%%%%%%%%%%%%%%%%%%%%%%%%%%%%%%%%%%%%%%
\subsection{Magic at the LHC and in charmonium decays}

In order to use quantum information observables we must first make sure that they are actually measurable  at colliders.  The recent analysis of the CMS Collaboration~\cite{CMS:2024pts} provides the values and uncertainties of the correlation coefficients for the production of top quark pairs:
\be
p \bar p \to t \bar t \, ,
\ee
at the LHC with $\sqrt{s} = 13$ TeV and an integrated luminosity of 138 fb$^{-1}$.
We take the central values of the correlation coefficients for $m_{t\bar t} > 800$ GeV and $|\cos \Theta | < 0.8$ and construct the corresponding density matrix (assuming vanishing polarizations). The concurrence is equal to
\be
\mathscr{C}  = 0.52 \pm 0.08\, ,
\ee
and therefore entanglement in the spins of the final state is significantly different from 0---in agreement with what was found in Ref.~\cite{CMS:2024pts}. The event bins considered in Ref.~\cite{CMS:2024pts} are not sufficiently high in invariant mass and not sufficiently narrow around $\cos \Theta=0$ to make possible a test of Bell nonlocality, according to the Monte Carlo simulation in Ref.~\cite{Fabbrichesi:2021npl} (see also Refs.~\cite{Han:2023fci,Cheng:2023qmz,Cheng:2024btk}).

We can use the same data to compute magic and find
\be
\mathscr{M}_{2}  = 0.54 \pm 0.06\, ,
\ee
which is the average value for events falling in the bin $m_{t\bar t} > 800$ GeV and $|\cos \Theta | < 0.8$.

Charmonium decays make it possible to test the presence of magic in the actual experimental data of other processes at high energy. Entanglement and Bell nonlocality have been shown to be present in these processes~\cite{Fabbrichesi:2024rec}.  As an example we take the process that is known with the least uncertainty, namely
\be
e^{+}e^{-} \to\gamma \to  c \bar c \to J/\psi \to \Lambda \bar \Lambda \, ,
\ee
in which the $J/\psi$ is produced polarized and decays into a pair of strange baryons. The spin correlation matrix of the two baryons depends on the scattering angle $\Theta$ because the polarization of the $J/\psi$ does.

Ten billion $J/\psi$ events have been collected at the BESIII detector~\cite{BESIII:2021cxx}. The decay $J/\psi \to \Lambda \bar \Lambda$ has branching fraction $(1.89 \pm 0.08) \times 10^{-3}$~\cite{Workman:2022ynf}. The decay into $\Lambda\bar \Lambda$ pairs is reconstructed from their dominant hadronic decays: $\Lambda \to p\pi^-$ and $\bar \Lambda \to \bar p \pi^-$. The maximum likelihood fit yields the values of the two parameters defining the helicity amplitudes~\cite{BESIII:2022qax}:  
  \be
 \alpha= 0.4748 \pm 0.0022 |_{\rm stat}\pm 0.0031|_{\rm syst} \quad \text{and} \quad \Delta \Phi= 0.7521\pm0.0042|_{\rm stat} \pm 0.0066|_{\rm syst}\, . \label{values}
 \ee 
No correlation in the uncertainties is provided.

 %%%%%%%%%%%%%%
\begin{figure}[h!]
\begin{center}
\includegraphics[width=3.3in]{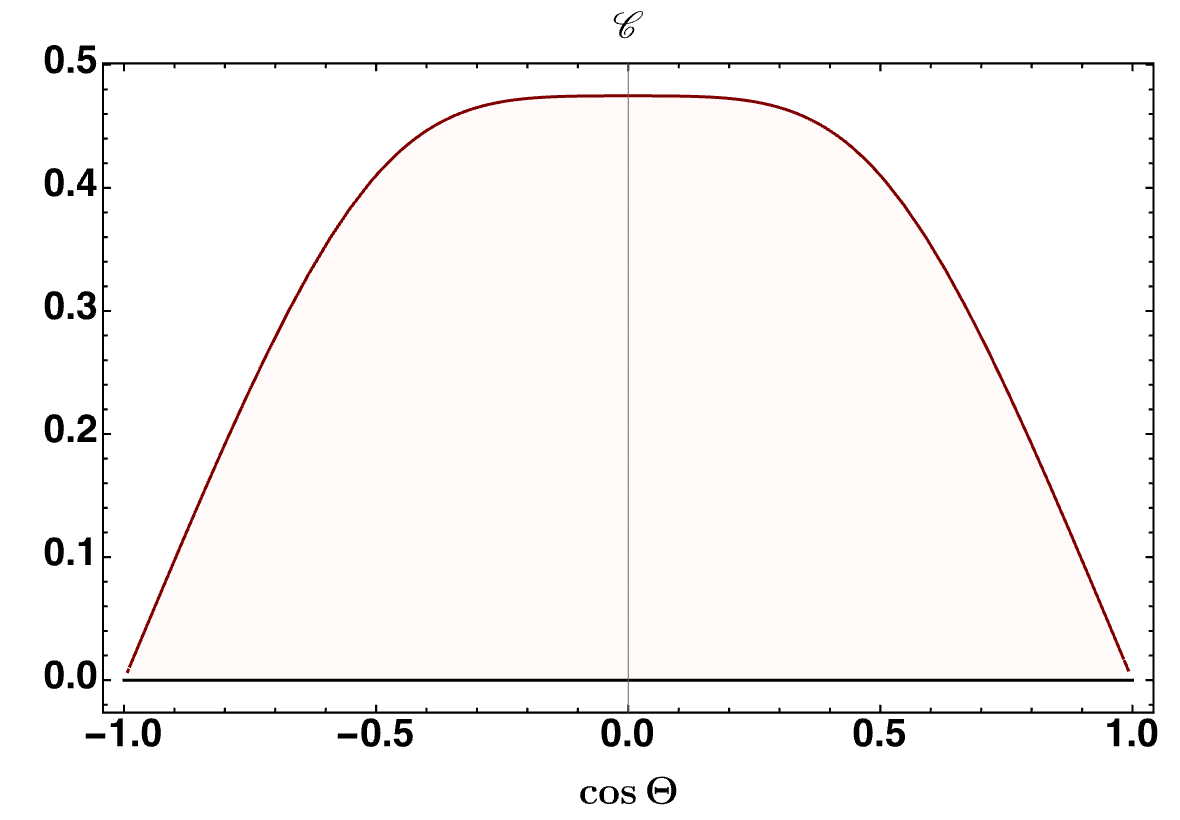}
\includegraphics[width=3.3in]{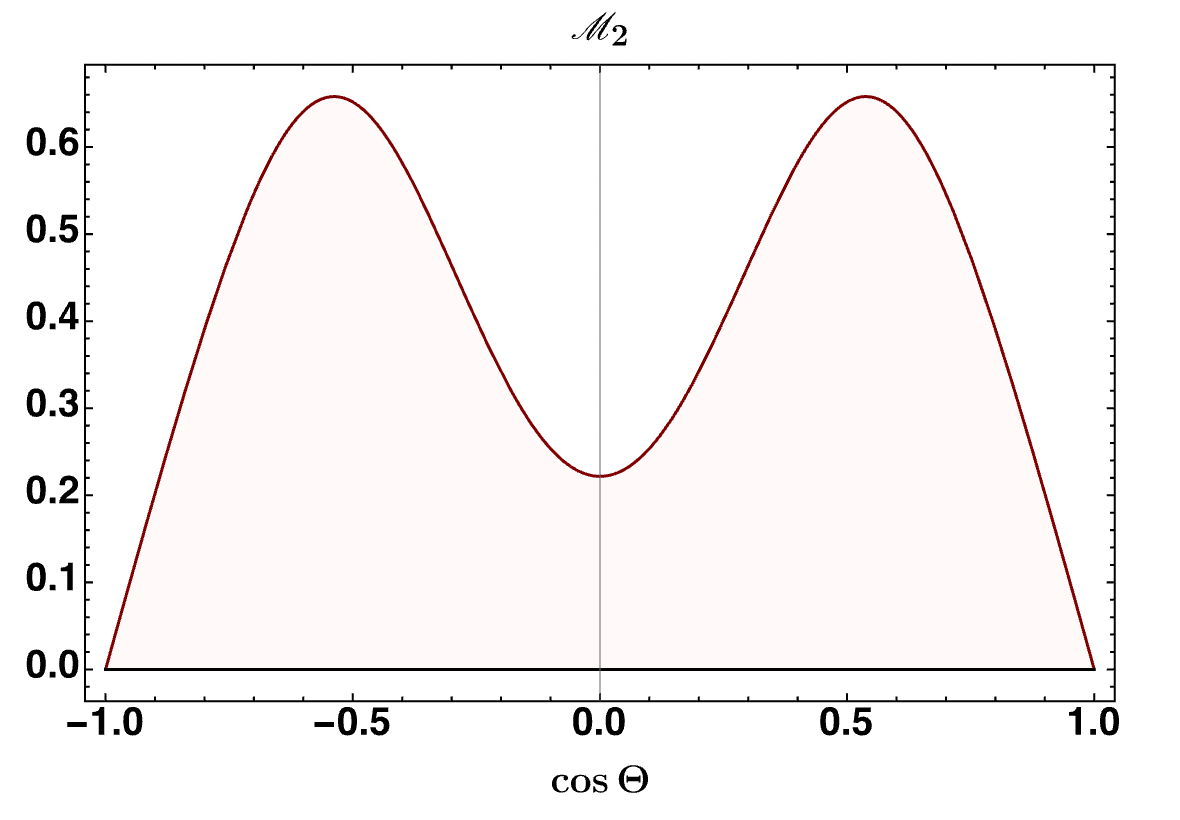}

\caption{\footnotesize 
\label{fig:magic} Concurrence (left panel) and magic (right panel) as a function of $\cos \Theta$ for the process $e^+ e^-\to J/\psi \to \Lambda \bar \Lambda$. }
\end{center}
\end{figure}
%%%%%%%%%%%%%%%%%%%%%%

The density matrix is given in \eq{eq:rhoHA} in terms of the helicity amplitudes. After writing the helicity amplitudes as 
 \be
 w_{\frac{1}{2}\, \frac{1}{2}}=w_{-\frac{1}{2}\, -\frac{1}{2}}=\frac{\sqrt{1-\alpha}}{\sqrt{2}}\quad \text{ and} \quad 
  w_{\frac{1}{2}\, -\frac{1}{2}} = w_{-\frac{1}{2}\, \frac{1}{2}}^*=i \sqrt{1+\alpha} \, \exp [-i \Delta \Phi ]\,,
  \ee
their experimental values are given by the numbers in \eq{values}.

The polarization of the $\Lambda$ baryons can then be computed from the density matrix in \eq{eq:rhoHA} through \eq{eq:FanoC}, obtaining
\be
\BB^-=\BB^+=\qty(-\frac{ \sqrt{1- \alpha^{2}} \sin 2 \, \Theta  \sin \Delta \Phi}{2 + \alpha + \alpha  \cos 2 \,\Theta},0, 0)\, . \label{pol}
\ee
Similarly, the spin correlation matrix is determined as:
\be
\CC = 
\frac{1}{C_{0}}\,
\begin{pmatrix}   
-2\,\alpha\,\sin^{2}\Theta  & 0 & 0 \\
0 & 2\,\sin^{2}\Theta & \sqrt{1-\alpha^2} \,\cos\Delta\Phi\,\sin2\Theta \\
0 &  \sqrt{1-\alpha^2} \,\cos\Delta\Phi\,\sin2\Theta & 1+2\,\alpha+\cos2\Theta
\end{pmatrix}\, , \label{cC}
\ee
in which $C_0=2 + \alpha + \alpha  \cos 2 \,\Theta$.

As shown in Fig.~\ref{fig:magic}, the concurrence and  magic depend on the scattering angle.
The largest values of magic are found at  $|\cos \Theta| = 0.53$, for which
\be
\mathscr{M}_{2}  = 0.658 \pm 0.002\, .
\ee
The presence of magic is established with a significance well above the $5\sigma$ level. 

%%%%%%%%%%%%%%%%%%%%%%%%%%%%%%%%%%%%%%%%%%%%%%%%%%%%%%%%%
\subsection{A comparison}

We compare the effectiveness of concurrence, magic, the trace distance, and the fidelity distance in constraining new physics by taking the case of the two anomalous couplings of the $\tau$ lepton, $F_2$ and $F_3$, at Belle and LEP3.

We use for the comparison a $\chi^2$ test given by
\begin{equation}
\chi^2(\lambda) = \qty(\frac{\mathcal{O}[\rho_{\text{NP}}(\lambda)]-\mathcal{O}(\rho_{\text{SM}})}{\sigma_\mathcal{O}})^2\leq (1.00)\; 3.84 \, ,
\end{equation}
set to highlight the (68\%) 95\% CL as we vary the new physics parameters $\lambda=F_2$, $F_3$ one at time. The uncertainty $\sigma_\mathcal{O}$ of each operator $\mathcal{O}=\mathscr{C}$, $\mathscr{M}_2$, $\mathscr{D}^T$, and $\mathscr{D}^F$ is computed by propagating a common set of uncertainties given for the Fano coefficients of the SM density matrix $\rho_{\text{SM}}=\rho_{\text{NP}}(\lambda=0)$. The involved density matrices are computed analytically for the CM energy of the considered experiments and averaged over the full range covered by the scattering angle. For the trace and fidelity distances it holds $\mathscr{D}^T(\rho_\text{SM})=\mathscr{D}^F(\rho_\text{SM})=0$.  

\begin{figure}[h!]
\begin{center}
\includegraphics[width=3.3in]{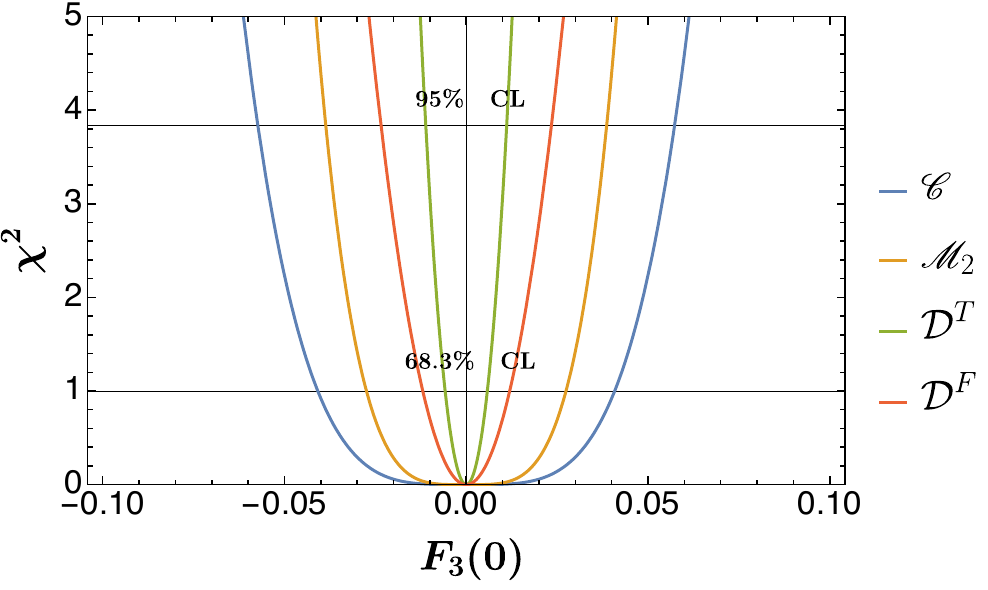}
\includegraphics[width=3.3in]{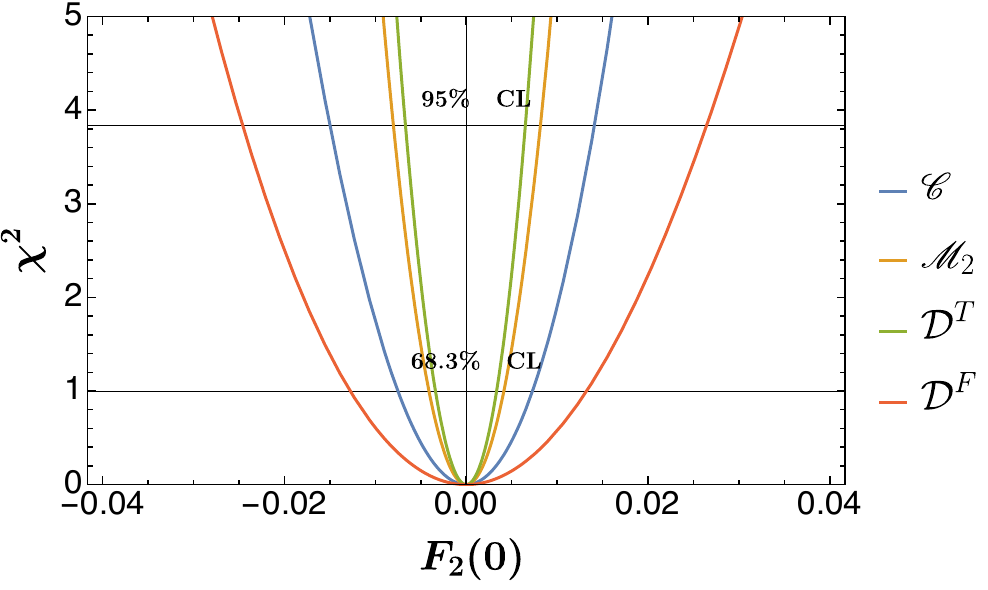}
\caption{\footnotesize 
\label{fig:compBelle} Comparison of $\chi^2$ tests obtained for $\mathscr{C}$, $\mathscr{M}_2$, $\mathscr{D}^T$, and $\mathscr{D}^F$ as we vary individually the $F_2$ and $F_3$ form factors of the $\tau$ lepton in the setup of the Belle experiment.
}
\end{center}
\end{figure}

\begin{figure}[h!]
\begin{center}
\includegraphics[width=3.3in]{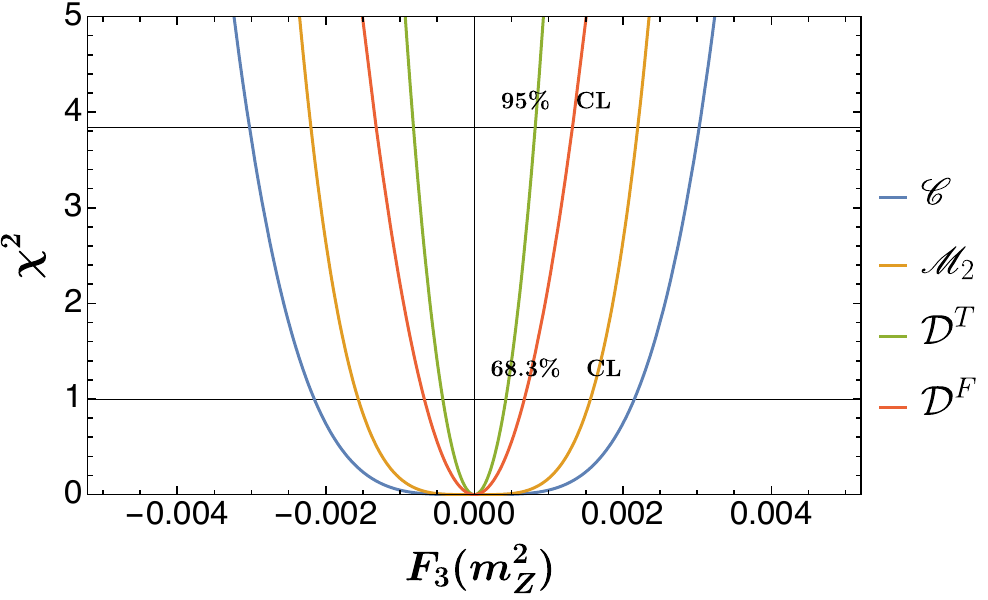}
\includegraphics[width=3.3in]{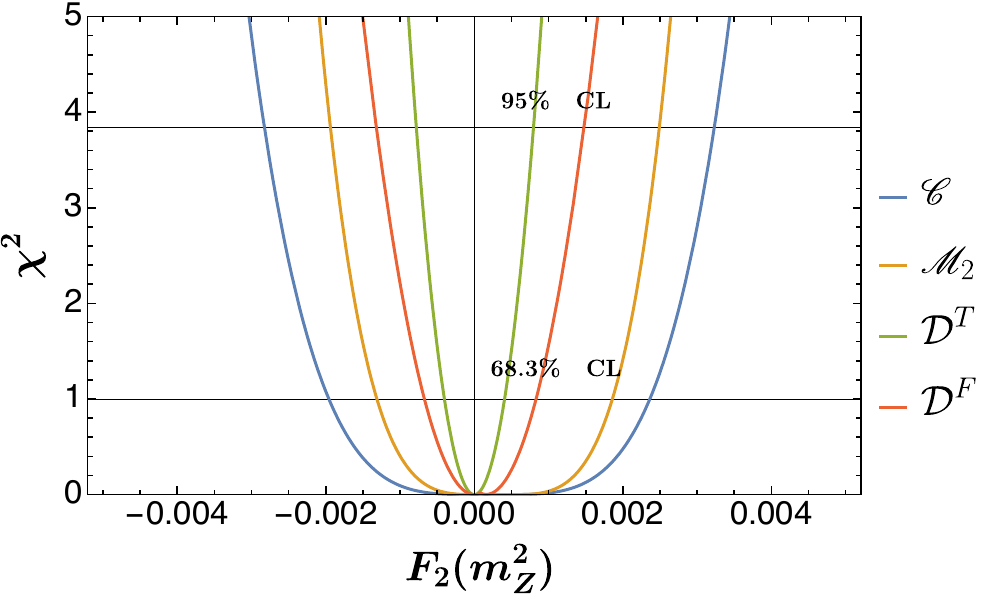}
\caption{\footnotesize 
\label{fig:compLEP3} Comparison of $\chi^2$ tests obtained for $\mathscr{C}$, $\mathscr{M}_2$, $\mathscr{D}^T$, and $\mathscr{D}^F$ as we vary individually the $F_2$ and $F_3$ form factors of the $\tau$ lepton in the setup of the LEP3 experiment.
}
\end{center}
\end{figure}

 The $\chi^2$ curves obtained for the different observables as we vary individually the sizes of the $F_2$ and $F_3$ form factors are shown in Fig.~\ref{fig:compBelle} and Fig.~\ref{fig:compLEP3}, respectively, for the Belle and LEP3 case.
%%%%%%%%%%%%%%
%%%%%%%%%%%%%%%%%%%%%%

%%%%%%%%%%%%%%
%%%%%%%%%%%%%%%%%%%%%%

We can see by inspection that the trace distance provides the most stringent limits, closely followed by the fidelity distance. This result holds for the two anomalous couplings here considered and for the CM energies used at the Belle and LEP3 experiments, as well as for the remaining anomalous couplings not shown here. The result lends support to our proposal to use the trace distance to highlight new physics effects in the spin correlation reconstructed via quantum tomography. 

The relative effectiveness of the other quantum information observables depends on the anomalous coupling and on the energy at which the limit is computed. Though the concurrence is better than magic for the limit on the anomalous magnetic moment testable at the Belle experiment, in general, it is less effective than magic in constraining the remaining anomalous couplings.

%%%%%%%%%%%%%%%%%%%%%%%%%%%%%%%%%%%%%%%%%%%%%%%%%%%%%%%%%
\section{Conclusions}
%%%%%%%%%%%%%%%%%%%%%%%%%%%%%%%%%%%%%%%%%%%%%%%%%%%%%%%%%

{\versal The inclusion of quantum information observables} in high-energy physics has introduced a variety of new variables into collider physics.  These new observables are sensitive to the spin information of final states and, consequently, are sensitive to new physics that alters this spin information.  While there have been studies that utilize some of these new variables in searches for new physics, there is no reason a priori that a particular quantum information observable is optimal for a particular new physics model.

This task of distinguishing between two hypotheses is studied in quantum information as similarity measures between density matrices.  In this work, we adopt these techniques, specifically the trace distance and the fidelity distance, into high-energy physics and apply them to several new physics models.  We show that in all cases, as expected, the trace distance outperforms other quantum information observables such as the concurrence or the magic.  The scenarios we study include a chromomagnetic top-quark dipole moment and anomalous couplings of the $\tau$ lepton.

The introduction of similarity measures into new physics searches naturally adapts the idea of globally fitting all relevant coefficients into quantities with well-defined quantum information properties.  While we have studied the trace distance and the fidelity distances, there are other distance measures that would be interesting to compare.  We are optimistic that distance measures will bridge the gap between the use quantum information at colliders and useful tools for new physics searches.

%%%%%%%%%%%%%%%%%%%%%%%%%%%%%%%%%%%%%%%%%%%%%%%%%%%%%%%%%%%%%%
%%%%%%%%%%%%%%%%%%%%%%%%%%%%%%%%%%%%%%%%%%%%%%%%%%%%%%%%%%%%%%
\vskip1cm
\section*{Acknowledgments}
{\small
ML is supported by the U.S.~Department of Energy under grant No.~DE-SC0007914 by the U.S.~National Science Foundation under grant No.~PHY-2112829. LM is supported by the Estonian Research Council under the RVTT3, PRG1884 and TK202 grants.
}

%%%%%%%%%%%%%%%%%%%%%%%%%%%%%%%%%%%%%%%%%%%%%%%%%%%%%%%%%%%%%%
%%%%%%%%%%%%%%%%%%%%%%%%%%%%%%%%%%%%%%%%%%%%%%%%%%%%%%%%%%%%%%

%%%%%%%%%%%%%%%%%%%%%%%%%%%%%%%%%%%%%%%%%%%%%%%%%%%%%%%%%%%%%%%
%%%%%%%%%%%%%%%%%%%%%%%%%%%%%%%%%%%%%%%%%%%%%%%%%%%%%%%%%%%%%%
\twocolumn  
\small
\bibliographystyle{JHEP}   
\bibliography{main.bib} 
\end{document}